\documentclass[journal,10pt,onecolumn,draftclsnofoot,]{IEEEtran}
\usepackage{graphicx}
\usepackage{epstopdf}
\usepackage{amssymb}
\usepackage{float}
\usepackage{xcolor}
\usepackage{hyperref}
\usepackage[linesnumbered,ruled]{algorithm2e}

\usepackage{caption}
\usepackage{subcaption}

\ifCLASSINFOpdf

\else

\fi

\usepackage{amsmath}

\begin{document}

\title{\LARGE Study of Robust Adaptive Power Allocation Techniques for Rate Splitting based MU-MIMO systems }
%
%
%

\author{A. R. Flores and R. C. de Lamare \vspace{-5.25em}
\thanks{This work was partially supported by the National Council for Scientific and Technological Development (CNPq), FAPERJ, FAPESP, and CGI.}\thanks{A. R. Flores is with the Centre for Telecommunications Studies, Pontifical
Catholic University of Rio de Janeiro, Rio de Janeiro 22451-900, Brazil (e-mail: andre.flores@cetuc.puc-rio.br).}\thanks{R. C. de Lamare is with the Centre for Telecommunications Studies,
Pontifical Catholic University of Rio de Janeiro, Rio de Janeiro 22451-900,
Brazil, and also with the Department of Electronic Engineering, University
of York, York YO10 5DD, U.K. (e-mail: delamare@cetuc.puc-rio.br).}}

\maketitle

\begin{abstract}
Rate splitting (RS) systems can better deal with imperfect channel state information at the transmitter (CSIT) than conventional approaches. However, this requires an appropriate power allocation that often has a high computational complexity, which might be inadequate for practical and large systems. To this end, adaptive power allocation techniques can provide good performance with low computational cost. This work presents novel robust and adaptive power allocation technique for RS-based multiuser multiple-input multiple-output (MU-MIMO) systems. In particular, we develop a robust adaptive power allocation based on stochastic gradient learning and the minimization of the mean-square error between the transmitted symbols of the RS system and the received signal. The proposed robust power allocation strategy incorporates knowledge of the variance of the channel errors to deal with imperfect CSIT and adjust power levels in the presence of uncertainty. An analysis of the convexity and stability of the proposed power allocation algorithms is provided, together with a study of their computational complexity and theoretical bounds relating the power allocation strategies. Numerical results show that the sum-rate of an RS system with adaptive power allocation outperforms RS and conventional MU-MIMO systems under imperfect CSIT. 
\end{abstract}
\begin{IEEEkeywords}
Rate splitting, multiuser MIMO, power allocation, ergodic sum rate.
\end{IEEEkeywords}

%
\IEEEpeerreviewmaketitle

\section{Introduction}
%
%
%
%

\IEEEPARstart{W}{ireless} communications systems employing multiple antennas have the advantage of increasing the overall throughput without increasing the required bandwidth. For this reason, multiple-antenna systems are at the core of several wireless communications standards such as WiFi, Long Term Evolution (LTE) and the fifth generation (5G). However, such wireless systems suffer from multiuser interference (MUI). In order to mitigate MUI, transmit processing techniques have been employed in the downlink (DL), allowing accurate recovery of the data at the receivers. In general, a precoding technique maps the symbols containing the information to the transmit antennas so that the information arrives at the receiver with reduced levels of MUI. Due to its benefits, linear \cite{mmimo,Joham2005,wence,grbd,wlrbd,rmmse} and non-linear \cite{Zu2014,Peel2005,bbprec} precoding techniques have been extensively reported in the literature. The design of effective precoders demands very accurate channel state information at the transmitter (CSIT), which is an extremely difficult task to accomplish in actual wireless systems. Hence, the transmitter typically only has access to partial or imperfect CSIT. As a result, the precoder cannot handle MUI as expected, resulting in residual MUI at the receiver. This residual MUI can degrade heavily the performance of wireless systems since it scales with the transmit power employed at the base station (BS) \cite{Tse2005}.

\subsection{Prior and related work}

In this context, Rate Splitting (RS) has emerged as a novel approach that is capable of dealing with CSIT imperfection \cite{Clerckx2016} in an effective way. RS was initially proposed in \cite{Han1981} to deal with interference channels \cite{Carleial1978}, where independent transmitters send information to independent receivers \cite{Haghi2021}.  Since then, several studies have found that RS outperforms conventional schemes such as conventional precoding in spatial division multiple access (SDMA), power-domain Non-Orthogonal Multiple Access (NOMA) \cite{Mao2018} and even dirty paper coding (DPC) \cite{Mao2020}. Interestingly, it turns out that RS constitutes a generalized framework which has as special cases other transmission schemes such SDMA, NOMA and multicasting \cite{Clerckx2020,Naser2020,Jaafar2020,Mao22}. The main advantage of RS is its capability to partially decode interference and partially treat interference as noise. To this end, RS splits the message of one or several users into a common message and a private message. The common message must be decoded by all the users that employ successive interference cancellation \cite{spa,mfsic,mbdf,bfidd,1bitidd}. On the other hand, the private messages are decoded only by their corresponding users.

RS schemes have been shown to enhance the performance of wireless communication systems. In \cite{Yang2013} RS was extended to the broadcast channel (BC) of multiple-input single-output (MISO) systems, where it was shown that RS provides gains in terms of Degrees-of-Freedom (DoF) with respect to conventional multiuser multiple-input multiple-output (MU-MIMO) schemes under imperfect CSIT. Later in \cite{Hao2017b}, the DoF of a MIMO BC and IC was characterized. RS has eventually been shown in \cite{Piovano2017} to achieve the optimal DoF region when considering imperfect CSIT, outperforming the DoF obtained by SDMA systems, which decays in the presence of imperfect CSIT.

Due to its benefits, several wireless communications deployments  with RS have been studied. RS has been employed in MISO systems along with linear precoders \cite{Joudeh2016,Hao2015} in order to maximize the sum-rate performance under perfect and imperfect CSIT assumptions.  Another approach has been presented in \cite{Joudeh2017} where the max-min fairness problem has been studied. In \cite{Hao2017} a K-cell MISO IC has been considered and the scheme known as topological RS presented. The topological RS scheme transmits multiple layers of common messages, so that  the common messages are not decoded by all users but by groups of users. RS has been employed  along with random vector quantization in \cite{Lu2018} to mitigate the effects of the imperfect CSIT caused by finite feedback. In \cite{Flores2020,rsthp}, RS with common stream combining techniques has been developed in order to exploit multiple antennas at the receiver and to improve the overall sum-rate performance. A successive null-space precoder, that employs null-space basis vectors to adjust the inter-user-interference at the receivers, is proposed in \cite{Krishnamoorthy2021}. The optimization of the precoders along with the transmission of multiple common streams was considered in \cite{Mishra2022}. In \cite{Li2020}, RS with joint decoding has been explored. The authors of \cite{Li2020} devised precoding algorithms for an arbitrary  number of users along with a stream selection strategy to reduce the number of precoded signals.

Along with the design of the precoders, power allocation is also a fundamental part of RS-based systems. The benefits of RS are obtained only if an appropriate power allocation for the common stream is performed. However, the power allocation problem in MU-MIMO systems is an NP-hard problem \cite{Luo2008,Liu2011}, and the optimal solution can be found at the expense of an exponential growth in computational complexity. Therefore, suboptimal approaches that jointly optimize the precoder and the power allocation have been developed. Most works so far employ exhaustive search or complex optimization frameworks. These frameworks rely on the augmented WMMSE \cite{Mao2020,Maoinpress,JoudehClerckx2016,Kaulich2021,Mishra2022}, which is an extension of the WMMSE proposed in \cite{Christensen2008}. This approach requires also an alternating optimization, which further increases the computational complexity. A simplified approach can be found in \cite{Dai2016a}, where closed-form expressions for RS-based massive MIMO systems are derived. However, this suboptimal solution is more appropiate for massive MIMO deployments. The high complexity of most optimal approaches makes them impractical to implement in large or real-time systems. For this reason, there is a strong demand for cost-effective power allocation techniques for RS systems.
\subsection{Contributions}

In this paper, we present novel efficient robust and adaptive power
allocation techniques \cite{rapa} for RS-based MU-MIMO systems. In
particular, we develop a robust adaptive power allocation (APA-R)
strategy based on stochastic gradient learning
\cite{bertsekas,jidf,smtvb,smce} and the minimization of the
mean-square error (MSE) between the transmitted common and private
symbols of the RS system and the received signal. We incorporate
knowledge of the variance of the channel errors to cope with
imperfect CSIT and adjust power levels in the presence of
uncertainty. When the knowledge of the variance of the channel
errors is not exploited the proposed APA-R becomes the proposed
adaptive power allocation algorithm (APA). An analysis of  the
convexity and stability of the proposed power allocation algorithms
along with a study of their computational complexity and theoretical
bounds relating the power allocation strategies are developed.
Numerical results show that the sum-rate of an RS system employing
adaptive power allocation outperforms conventional MU-MIMO systems
under imperfect CSIT assumption. The contributions of this work can
be summarized as:
\begin{itemize}
    \item Cost-effective APA-R and APA techniques for power allocation are proposed based on stochastic gradient recursions and knowledge of the variance of the channel errors for RS-based and standard MU-MIMO systems.
    \item An analysis of convexity and stability of the proposed power allocation techniques along with a bound on the MSE of APA and APA-R and a study of their computational complexity.
    \item A simulation study of the proposed APA and APA-R, and the existing power allocation techniques for RS-based and standard MU-MIMO systems.
\end{itemize}

\subsection{Organization}
The rest of this paper is organized as follows. Section II describes the mathematical model of an RS-based MU-MIMO system. In Section III the proposed APA-R technique is presented, the proposed APA approach is obtained as a particular case and sum-rate expressions are derived. In Section IV, we present an analysis of convexity and stability of the proposed APA and APA-R techniques along with a bound on the MSE of APA and APA-R and a study of their computational complexity. Simulation results are illustrated and discussed in Section V. Finally, Section VI presents the conclusions of this work.

\subsection{Notation}
Column vectors are denoted by lowercase boldface letters. The vector $\mathbf{a}_{i,*}$ stands for the $i$th row of matrix $\mathbf{A}$. Matrices are denoted by uppercase boldface letters. Scalars are denoted by standard letters. The superscript $\left(\cdot\right)^{\text{T}}$ represents the transpose of a matrix, whereas the notation $\left(\cdot\right)^H$ stands for the conjugate transpose of a matrix. The operators $\lVert \cdot \rVert$, and $\mathbb{E}_x\left[\cdot\right]$ represent the Euclidean norm, and the expectation w.r.t the random variable $x$, respectively. The trace operator is given by $\text{tr}\left(\cdot\right)$.  The Hadamard product is denoted by $\odot$. The operator $\text{diag}\left(\mathbf{a}\right)$ produces a diagonal matrix with the coefficients of $\mathbf{a}$ located in the main diagonal.

\section{System Model}

\begin{figure}[htb!]
\begin{center}
\includegraphics[scale=0.45]{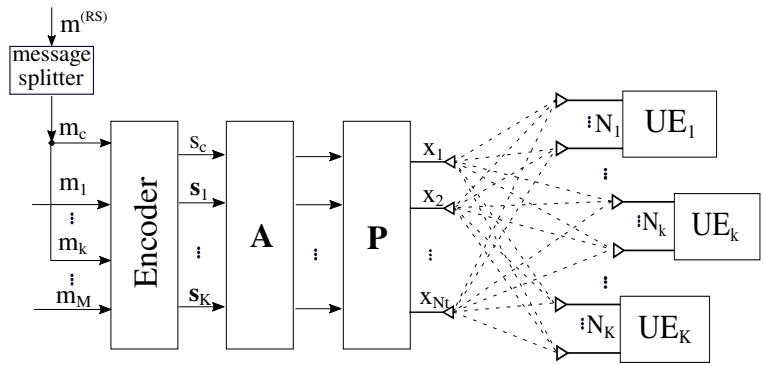}
\vspace{-1.0em}
\caption{RS MU-MIMO architecture.}
\label{System Model RS}
\end{center}
\vspace{-2em}
\end{figure}

Let us consider the RS-based MU-MIMO system architecture depicted in Fig. \ref{System Model RS}, where the BS is equipped with $N_t$ antennas, serves $K$ users and the $k$th UE is equipped with $N_k$ antennas. Let us denote by $N_r$ the total number of receive antennas. Then, $N_r=\sum_{k=1}^K N_k$. For simplicity, the message intended for the $k$th user is split into a common message and a private message. Then, the messages are encoded and modulated. The transmitter sends one common stream and a total of $M$ private streams, with $M\leq N_r$. The set $\mathcal{M}_k$  contains the $M_k$ private streams, that are intended for the user $k$, where $M_k\leq N_k$. It follows that $M=\sum_{k^=1}^K M_k$.

The vector $\mathbf{s}^{\left(\text{RS}\right)}=\left[s_c,\mathbf{s}_p^{\text{T}}\right]^{\text{T}} \in \mathbb{C}^{M+1}$, which is assumed i.i.d. with zero mean and covariance matrix equal to the identity matrix, contains the information transmitted to all users, where $s_c$ is the common symbol and $\mathbf{s}_p=\left[\mathbf{s}_1^{\text{T}},\mathbf{s}_2^{\text{T}},\cdots,\mathbf{s}_K^{\text{T}}\right]^{\text{T}}$  contains the private symbols of all users. Specifically, the vector $\mathbf{s}_k \in \mathbb{C}^{M_k}$ contains the private streams intended for the $k$th user. The system is subject to a transmit power constraint given by $\mathbb{E}\left[\lVert\mathbf{x}^{\left(\text{RS}\right)}\rVert^2\right]\leq E_{tr}$, where $\mathbf{x}^{\left(\text{RS}\right)}\in \mathbb{C}^{N_t}$ is the transmitted vector and $E_{tr}$ denotes the total available power. The transmitted vector can be expressed as follows:
\begin{align}
\mathbf{x}^{\left(\text{RS}\right)}=&\mathbf{P}^{\left(\text{RS}\right)}\mathbf{A}^{\left(\text{RS}\right)}\mathbf{s}^{\left(\text{RS}\right)}=a_c s_c \mathbf{p}_c+\sum_{m=1}^{M}a_m s_m \mathbf{p}_m, \label{Transmit Signal}
\end{align}
where $\mathbf{A}^{\left(\text{RS}\right)}\in \mathbb{R}^{\left(M+1\right)\times \left(M+1\right)}$ represents the power allocation matrix and $\mathbf{P}^{\left(\text{RS}\right)}=[\mathbf{p}_c,\mathbf{p}_1,\cdots ,\mathbf{p}_K] \in \mathbb{C}^{N_t \times \left(M+1\right)}$ is used to precode the vector of symbols $\mathbf{s}^{\left(\text{RS}\right)}$. Specifically, $\mathbf{A}^{\text{RS}}=\text{diag}\left(\mathbf{a}^{\left(\text{RS}\right)}\right)$ and $\mathbf{a}^{\left(\text{RS}\right)}=\left[a_c, a_1,\cdots,a_M\right]^{\text{T}}\in \mathbb{R}^{M+1}$, where $a_c$ denotes the power allocated to the common stream and $a_k$ allocates power to the $k$th private stream. Without loss of generality, we assume that the columns of the precoders are normalized to have unit norm.

The model established leads us to the received signal described by
\begin{equation}
\mathbf{y}=\mathbf{H}\mathbf{P}^{\left(\text{RS}\right)}\text{diag}\left(\mathbf{a}^{\left(\text{RS}\right)}\right)\mathbf{s}^{\left(\text{RS}\right)}+\mathbf{n}, \label{Receive Signal Complete}
\end{equation}
where $\mathbf{n}=\left[\mathbf{n}_1^{\text{T}},\mathbf{n}_2^{\text{T}},\cdots,\mathbf{n}_K^{\text{T}}\right]^{\text{T}} \in \mathbb{C}^{N_r}$ represents the uncorrelated noise vector, which follows a complex normal distribution, i.e., $\mathbf{n}\sim \mathcal{CN}\left(\mathbf{0},\sigma_n^2\mathbf{I}\right)$. We assume that the noise and the symbols are uncorrelated, which is usually the case in real systems. The matrix $\mathbf{H}=\left[\mathbf{H}_1^{\text{T}},\mathbf{H}_2^{\text{T}},\cdots,\mathbf{H}_K^{\text{T}}\right]^{\text{T}}\in \mathbb{C}^{N_r\times N_t}$ denotes the channel between the BS and the user terminals. Specifically, $\mathbf{n}_k$ denotes the noise affecting the $k$th user and the matrix $\mathbf{H}_k\in \mathbb{C}^{N_k\times N_t}$ represents the channel between the BS and the $k$th user. The imperfections in the channel estimate are modelled by the random matrix $\tilde{\mathbf{H}}$. Each coefficient of $\tilde{\mathbf{H}}$ follows a Gaussian distribution with variance equal to $\sigma_{e,i}^2$ and $\mathbb{E}\left[\tilde{\mathbf{h}}_{i,*}^H\tilde{\mathbf{h}}_{i,*}\right]=\sigma_e^2\mathbf{I}\quad \forall i \in\left[1,N_r\right]$. Then, the channel matrix can be expressed as $\mathbf{H}=\hat{\mathbf{H}}+\tilde{\mathbf{H}}$, where the channel estimate is given by $\hat{\mathbf{H}}$. From \eqref{Receive Signal Complete} we can obtain the received signal of user $k$, which is given by
\begin{align}
\mathbf{y}_k=&a_c s_c \mathbf{H}_k\mathbf{p}_c+ \mathbf{H}_k\sum_{i\in \mathcal{M}_k}a_i s_i\mathbf{p}_i+\mathbf{H}_k\sum\limits_{\substack{l=1\\l \neq k}}^{K}\sum\limits_{j\in \mathcal{M}_l}a_j s_j\mathbf{p}_j+ \mathbf{n}_k.\label{Receive Signal per user}
\end{align}
Note that the RS architecture contains the conventional MU-MIMO as a special case where no message is split and therefore $a_c$ is set to zero. Then, the model boils down to the model of a conventional MU-MIMO system, where the received signal at the $k$th user is given by
\begin{equation}
    \mathbf{y}_k=\mathbf{H}_k\sum_{i\in \mathcal{M}_k}a_i s_i\mathbf{p}_i+\mathbf{H}_k\sum\limits_{\substack{l=1\\l \neq k}}^{K}\sum\limits_{j\in \mathcal{M}_l}a_j s_j\mathbf{p}_j+\mathbf{n}_k\label{Receive Signal per user convetinoal MIMO}
\end{equation}
In what follows, we will focus on the development of power allocation techniques that can cost-effectively compute $a_c$ and  $a_j, j = 1, \ldots, K$.

\section{Proposed Power Allocation Techniques}

In this section, we detail the proposed power allocation techniques. In particular, we start with the derivation of the ARA-R approach and then present the APA technique as a particular case of the APA-R approach.

\subsection{Robust Adaptive Power Allocation}\label{c5 section robust power allocation RS}

Here, a robust adaptive power allocation algorithm, denoted as APA-R, is developed to perform power allocation in the presence of imperfect CSIT. The key idea is to incorporate knowledge about the variance of the channel uncertainty \cite{locsme,okspme} into an adaptive recursion to allocate the power among the streams. The minimization of the MSE between the received signal and the transmitted symbols is adopted as the criterion to derive the algorithm.

Let us consider the model established in \eqref{Receive Signal Complete} and denote the fraction of power allocated to the common stream by the parameter $\delta$, i.e., $a_c^2=\delta E_{tr}$. It follows that the available power for the private streams is reduced to $\left(1-\delta\right)E_{tr}$. We remark that the length of $\mathbf{s}^{\left(\text{RS}\right)}$ is greater than that of $\mathbf{y}^{\left(\text{RS}\right)}$ since the common symbol is superimposed to the private symbols.  Therefore, we consider the vector $\mathbf{y}'=\mathbf{T}\mathbf{y}^{\left(\text{RS}\right)}$, where $\mathbf{T}\in\mathbb{R}^{\left(M+1\right)\times M}$ is a transformation matrix employed to ensure that the dimensions of $\mathbf{s}^{\left(\text{RS}\right)}$ and $\mathbf{y}^{\left(\text{RS}\right)}$ match, and is given by
\begin{equation}
    \mathbf{T}=\begin{bmatrix}
    1 &1 &\cdots &1\\
    1 &0 &\cdots &0\\
    0 &1 &\cdots &0\\
    \vdots &\vdots &\ddots &\vdots\\
    0 &0 &\cdots &1
    \end{bmatrix}.
\end{equation}
All the elements in the first row of matrix $\mathbf{T}$ are equal to one in order to take into account the common symbol obtained at all receivers. As a result we obtain the combined receive signal of all users. It follows that
\begin{equation}
    \mathbf{y}'=\begin{bmatrix}
    y_c\\
    y_1\\
    \vdots\\
    y_M
    \end{bmatrix}=\begin{bmatrix}
    \sum_{i=1}^{M} y_i\\
    y_1\\
    \vdots\\
    y_M
    \end{bmatrix},
\end{equation}
where the received signal at the $i$th antenna is described by
\begin{equation}
        y_i=a_c s_c\left(\hat{\mathbf{h}}_{i,*}+\tilde{\mathbf{h}}_{i,*}\right)\mathbf{p}_c+\sum_{j=1}^{M}a_j s_j \left(\hat{\mathbf{h}}_{i,*}+\tilde{\mathbf{h}}_{i,*}\right)\mathbf{p}_j+ n_i.
\end{equation}

Let us now consider the proposed robust power allocation problem for imperfect CSIT scenarios. By including the error of the channel estimate, the robust power allocation problem can be formulated as the constrained optimization given by

\begin{equation}
\begin{gathered}
\min_{\mathbf{a}} \mathbb{E}\left[\lVert\mathbf{s}^{\left(\text{RS}\right)}-\mathbf{y}'\left(\mathbf{H}\right)\rVert^2| \hat{\mathbf{H}}\right]\\
\text{s.t.}~~\text{tr}\left(\mathbf{P}^{\left(\text{RS}\right)}\text{diag}\left(\mathbf{a}^{\left(\text{RS}\right)}\odot \mathbf{a}^{\left(\text{RS}\right)}\right)\mathbf{P}^{\left(\text{RS}\right)^{H}}\right)=E_{tr},\label{Obejctive funtion robust}
\end{gathered}
\end{equation}
which can be solved by first relaxing the constraint, using an adaptive learning recursion and then enforcing the constraint.

\begin{figure}
    \centering
    \includegraphics[scale=0.4]{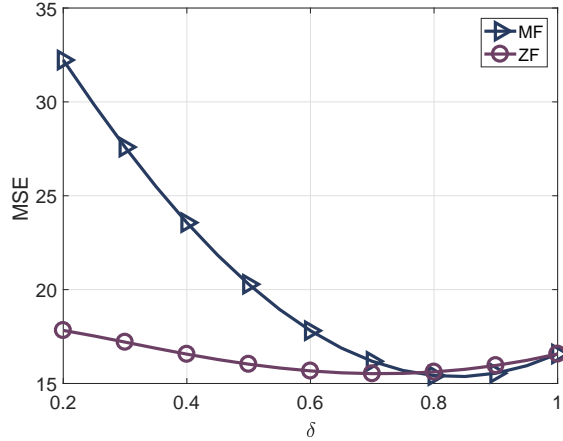}
    \vspace{-1.5em}\caption{Objective function considering a MU-MIMO system with $Nt=3$, $K=3$, and $\sigma_n^2=1$}\vspace{-1.5em}
    \label{fig:MSECurve}
\end{figure}

In this work, we choose the MSE as the objective function due to its convex property and mathematical tractability, which help to find an appropriate solution through algorithms. The objective function is convex as illustrated by Fig. \ref{fig:MSECurve} and analytically shown in Section \ref{analysis}. In Fig. \ref{fig:MSECurve}, we plot the objective function using two precoders, namely the zero-forcing (ZF) and the matched filter (MF) precoders \cite{Joham2005}, where three private streams and one common stream are transmitted and the parameter $\delta$ varies with uniform power allocation across private streams.

To solve \eqref{Obejctive funtion robust} we need to expand the terms and evaluate the expected value. Let us consider that the square error is equal to $\varepsilon=\lVert \mathbf{s}^{\left(\text{RS}\right)}-\mathbf{y}' \rVert^2$. Then, the MSE is given by
\begin{align}
    \mathbb{E}\left[\varepsilon|\hat{\mathbf{H}}\right]=&-2a_c\sum_{i=1}^M\Re{\left\{\hat{\phi}^{\left(i,c\right)}\right\}}-2\sum_{j=1}^M a_j \Re{\left\{\hat{\phi}^{\left(j,j\right)}\right\}}+2\sum_{j=1}^{M}a_j^2\left(\sum_{l=1}^M\lvert\hat{\phi}^{\left(l,j\right)}\rvert^2+M\sigma_{e,i}^2\lVert\mathbf{p}_j\rVert^2\right)\nonumber\\
    &+2a_c^2\left(\sum_{i=1}^M\lvert\hat{\phi}^{\left(i,c\right)}\rvert^2+M\sigma_{e,i}^2\lVert\mathbf{p}_c\rVert^2\right)+2\sum_{i=1}^{M-1}\sum_{q=i+1}^{M}\sum_{r=1}^M a_r^2\Re\left\{\hat{\phi}^{\left(i,r\right)^*}\hat{\phi}^{\left(q,r\right)}\right\}\nonumber\\
        &+\sum_{l=1}^M\sum\limits_{\substack{j=1\\j\neq l}}^M a_c^2\hat{\phi}^{\left(l,c\right)^*}\hat{\phi}^{\left(j,c\right)}+M\left(1+2\sigma_n^2\right)+1,\label{mean square error APA robust}
\end{align}
where  $\hat{\phi}^{\left(i,c\right)}=\hat{\mathbf{h}}_{l,*}\mathbf{p}_c$ and $\hat{\phi}^{\left(i,l\right)}=\hat{\mathbf{h}}_{i,*}\mathbf{p}_l$ for all $i,l \in \left[1,M\right]$. The proof to obtain the last equation can be found in appendix \ref{Appendix MSE APA-R}. The partial derivatives of \eqref{mean square error APA robust} with respect to ${a}_c$ and ${a}_i$ are expressed by
\begin{align}
 \frac{\partial\mathbb{E}\left[\varepsilon|\hat{\mathbf{H}}\right]}{\partial a_c}=& 2\sum_{l=1}^M\sum\limits_{\substack{j=1\\j\neq l}}^M a_c\hat{\phi}^{\left(l,c\right)^*}\hat{\phi}^{\left(j,c\right)}-2\sum_{i=1}^M\Re{\left\{\hat{\phi}^{\left(i,c\right)}\right\}}+4a_c\left(\sum_{i=1}^M\lvert\hat{\phi}^{\left(i,c\right)}\rvert^2+M\sigma_{e,i}^2\lVert\mathbf{p}_c\rVert^2\right),\label{gradient robust apa ac}
\end{align}
\begin{align}
    \frac{\partial\mathbb{E}\left[\varepsilon|\hat{\mathbf{H}}\right]}{\partial a_i}=&4a_i\sum_{l=1}^{M-1}\sum_{q=l+1}^{M} \Re\left[\hat{\phi}^{\left(l,i\right)^*}\hat{\phi}^{\left(q,i\right)}\right]-2 \Re{\left\{\hat{\phi}^{\left(i,i\right)}\right\}}+4a_i\left(\sum_{l=1}^M\lvert\hat{\phi}^{\left(l,i\right)}\rvert^2+M\sigma_{e,i}^2\lVert\mathbf{p}_i\rVert^2\right).\label{gradient robust apa ai}
\end{align}
 The partial derivatives given by \eqref{gradient robust apa ac} and \eqref{gradient robust apa ai} represent the gradient of the MSE with respect to the power allocation coefficients. With the obtained gradient we can employ a gradient descent algorithm, which is an iterative procedure that finds a local minimum of a differentiable function. The key idea is to take small steps in the opposite direction of the gradient, since this is the direction of the steepest descent. Remark that the objective function used is convex and has no local minimum. Then, the recursions of the proposed APA-R technique are given by
 \begin{align}
     a_c\left[t+1\right]&=a_c\left[t\right]-\mu\frac{\partial\mathbb{E}\left[\lvert\varepsilon\rvert^2|\hat{\mathbf{H}}^{\text{T}}\right]}{\partial a_c},\nonumber\\
     a_i\left[t+1\right]&=a_i\left[t\right]-\mu\frac{\partial\mathbb{E}\left[\lvert\varepsilon\rvert^2|\hat{\mathbf{H}}^{\text{T}}\right]}{\partial a_i},
 \end{align}
where the parameter $\mu$ represents the learning rate of the adaptive algorithm. At each iteration, the power constraint is analyzed. Then, the coefficients are scaled with a power scaling factor by $\mathbf{a}^{\left(\rm{RS}\right)}\left[n\right]=\beta\mathbf{a}^{\left(\rm{RS}\right)}\left[n\right]$, where $\beta=\sqrt{\frac{1}{\textrm{tr}\left(\textrm{diag}\left(\mathbf{a}^{\left(\rm{RS}\right)}\left[n\right]\odot \mathbf{a}^{\left(\rm{RS}\right)}\left[n\right]\right)\right)}}$ to ensure that the power constraint is satisfied. Algorithm \ref{algorithm RS APA} summarizes the proposed APA-R algorithm.

 \begin{algorithm}[t!]
 \normalsize
\SetAlgoLined
 given $\hat{\mathbf{H}}$, $\mathbf{P}$, and $\mu$\;
 $\mathbf{a}\left[1\right]=\mathbf{0}$\;
 \For{$n=2$ \KwTo $I_t$}{
  $\frac{\partial\mathbb{E}\left[\varepsilon|\hat{\mathbf{H}}\right]}{\partial a_c}= 2\sum\limits_{l=1}^M\sum\limits_{\substack{j=1\\j\neq l}}^M a_c\hat{\phi}^{\left(l,c\right)^*}\hat{\phi}^{\left(j,c\right)}-2\sum\limits_{i=1}^M\Re{\left\{\hat{\phi}^{\left(i,c\right)}\right\}}+4a_c\left(\sum\limits_{i=1}^M\lvert\hat{\phi}^{\left(i,c\right)}\rvert^2+M\sigma_{e,i}^2\lVert\mathbf{p}_c\rVert^2\right)$\;
   \vspace{0.1cm}
  $\frac{\partial\mathbb{E}\left[\varepsilon|\hat{\mathbf{H}}\right]}{\partial a_i}=4a_i\sum\limits_{l=1}^{M-1}\sum\limits_{q=l+1}^{M} \Re\left[\hat{\phi}^{\left(l,i\right)^*}\hat{\phi}^{\left(q,i\right)}\right]-2 \Re{\left\{\hat{\phi}^{\left(i,i\right)}\right\}}+4a_i\left(\sum\limits_{l=1}^M\lvert\hat{\phi}^{\left(l,i\right)}\rvert^2+M\sigma_{e,i}^2\lVert\mathbf{p}_i\rVert^2\right)$\;
  \vspace{0.1cm}
  $a_c\left[n\right]=a_c\left[n-1\right]-\mu\frac{\partial\mathbb{E}\left[\varepsilon|\hat{\mathbf{H}}\right]}{\partial a_c}$\;
    \vspace{0.1cm}
    $a_i\left[n\right]=a_i\left[n-1\right]-\mu\frac{\partial\mathbb{E}\left[\varepsilon|\hat{\mathbf{H}}\right]}{\partial a_i}$\;
    \vspace{0.1cm}
  \If{$\textrm{\rm tr}\left(\textrm{\rm diag}\left(\mathbf{a}^{\left(\rm{RS}\right)}\left[n\right]\odot \mathbf{a}^{\left(\rm{RS}\right)}\left[n\right]\right)\right)\neq 1$}{
     \vspace{0.1cm}
   $\beta=\sqrt{\frac{1}{\textrm{tr}\left(\textrm{diag}\left(\mathbf{a}^{\left(\rm{RS}\right)}\left[n\right]\odot \mathbf{a}^{\left(\rm{RS}\right)}\left[n\right]\right)\right)}}$\;
      \vspace{0.1cm}
   $\mathbf{a}^{\left(\rm{RS}\right)}\left[n\right]=\beta\mathbf{a}^{\left(\rm{RS}\right)}\left[n\right]$\;
   }
 }
 \caption{Robust Adaptive Power allocation}
 \label{algorithm RS APA}
\end{algorithm}\vspace{-0.5em}

\subsection{Adaptive Power Allocation}

In this section, a simplified version of the proposed APA-R algorithm is derived. The main objective is to reduce the complexity of each recursion of the adaptive algorithm and avoid the load of computing the statistical parameters of the imperfect CSIT, while reaping the benefits of RS systems.
The power allocation problem can be reformulated as the constrained optimization problem given by
\begin{equation}
\begin{gathered}
\min_{\mathbf{a}} \mathbb{E}\left[\lVert\mathbf{s}^{\left(\text{RS}\right)}-\mathbf{y}'\rVert^2\right]\\
\text{s.t.}~~\text{tr}\left(\mathbf{P}^{\left(\text{RS}\right)}\text{diag}\left(\mathbf{a}^{\left(\text{RS}\right)}\odot \mathbf{a}^{\left(\text{RS}\right)}\right)\mathbf{P}^{\left(\text{RS}\right)^{H}}\right)=E_{tr},
\end{gathered} \label{Objective function apa}
\end{equation}
In this case, the MSE is equivalent to
\begin{align}
     \mathbb{E}\left[\varepsilon\right]=&-2a_c\sum_{j=1}^{M}\Re\left\{\phi^{\left(j,c\right)}\right\}-2\sum_{l=1}^M a_l\Re\left\{\phi^{\left(l,l\right)}\right\}+2\left(\sum_{l=1}^M a_c^2\lvert\phi^{\left(l,c\right)}\rvert^2+\sum_{i=1}^M\sum_{j=1}^M a_j^2\lvert\phi^{\left(i,j\right)}\rvert^2\right)\nonumber\\
     &+2\sum_{i=1}^{M-1}\sum_{q=i+1}^{M}\sum_{r=1}^M a_r^2\Re\left\{\phi^{\left(i,r\right)^*}\phi^{\left(q,r\right)}\right\}+\sum_{l=1}^M\sum\limits_{\substack{j=1\\j\neq l}}^M a_c^2\phi^{\left(l,c\right)^*}\phi^{\left(j,c\right)}+M\left(1+2\sigma_n^2\right)+1, \label{mean square error APA RS}
     \end{align}
 where we considered that $\phi^{\left(i,c\right)}=\mathbf{h}_{l,*}\mathbf{p}_c$ and $\phi^{\left(i,l\right)}=\mathbf{h}_{i,*}\mathbf{p}_l$ for all $i,l \in \left[1,M\right]$. The proof to obtain \eqref{mean square error APA RS} can be found in  appendix \ref{Appendix MSE APA}. Taking the partial derivatives of \eqref{mean square error APA RS} with respect to the coefficients of  $\mathbf{a}^{\left(\text{RS}\right)}$ we arrive at
 \begin{align}
     \frac{\partial\mathbb{E}\left[\varepsilon\right]}{\partial a_c}&=4a_c \sum_{i=1}^{M}\lvert\phi^{\left(i,c\right)}\rvert^2+2a_c\sum_{l=1}^M\sum\limits_{\substack{j=1\\j\neq l}}^M \phi^{\left(l,c\right)^*}\phi^{\left(j,c\right)}-2\sum_{q=1}^{M}\Re\left[\phi^{\left(q,c\right)}\right],\label{gradient RS perfect CSIT common stream}\\
     \frac{\partial\mathbb{E}\left[\varepsilon\right]}{\partial a_i}&=4a_i\sum_{j=1}^M \lvert\phi^{\left(j,i\right)}\rvert^2+4a_i\sum_{r=1}^{M-1}\sum_{q=r+1}^{M} \Re\left[\phi^{\left(r,i\right)^*}\phi^{\left(q,i\right)}\right]-2 a_i\Re\left[\phi^{\left(i,i\right)}\right],\label{gradient RS perfect CSIT private streams}
 \end{align}
  The power allocation coefficients are adapted using \eqref{gradient RS perfect CSIT common stream} and \eqref{gradient RS perfect CSIT private streams} in the following recursions:
 \begin{align}
     a_c\left[t+1\right]&=a_c\left[t\right]-\mu\frac{\partial\mathbb{E}\left[\varepsilon\right]}{\partial a_c}\nonumber\\
     a_i\left[t+1\right]&=a_i\left[t\right]-\mu\frac{\partial\mathbb{E}\left[\varepsilon\right]}{\partial a_i}.\label{update equation for perfect CSIT}
 \end{align}

\subsection{Sum-Rate Performance}
In this section, we derive closed-form expressions to compute the sum-rate performance of the proposed algorithms. Specifically, we employ the ergodic sum-rate (ESR) as the main performance metric. Before the computation of the ESR we need to find the average power of the received signal, which is given by
\begin{equation}
    \mathbb{E}\left[\lvert y_k\rvert^2\right]=a_c^2\lvert \mathbf{h}_k^{\textrm{T}}\mathbf{p}_c\rvert^2+\sum_{i=1}^K a_i^2\lvert \mathbf{h}_k^{\textrm{T}}\mathbf{p}_i\rvert^2+\sigma_w^2.
\end{equation}
It follows that the instantaneous SINR while decoding the common symbol is given by
\begin{align}
    \gamma_{c,k}&=\frac{a_c^2\lvert \mathbf{\hat{h}}_k^{\textrm{T}}\mathbf{p}_c\rvert^2}{\sum\limits_{i=1}^K a_i^2\lvert \mathbf{h}_k^{\textrm{T}}\mathbf{p}_i\rvert^2+\sigma_w^2}.\label{instantaneous SINR common rate}
\end{align}
 Once the common symbol is decoded, we apply SIC to remove it from the received signal. Afterwards, we calculate the instantaneous SINR when decoding the $k$th private stream, which is given by
\begin{equation}
    \gamma_k=\frac{a_k^2\lvert\mathbf{\hat{h}}_k^{\textrm{T}}\mathbf{p}_k\rvert^2}{\sum\limits_{\substack{i=1\\i\neq k}}^K a_i^2\lvert\mathbf{h}_k\mathbf{p}_i\rvert^2+\sigma_w^2}.\label{instantaneous SINR private rate}
\end{equation}
 Considering Gaussian signaling, the instantaneous common rate can be found with the following equation:
\begin{equation}
     R_{c,k}=\log_2\left(1+\gamma_{c,k}\right).\label{instantaneous common rate per user}
\end{equation}
The private rate of the $k$th stream is given by
\begin{equation}
     R_{k}=\log_2\left(1+\gamma_{k}\right)\label{instantaneous private rate}
\end{equation}
Since imperfect CSIT is being considered, the instantaneous rates are not achievable. To that end, we employ the average sum rate (ASR) to average out the effect of the error in the channel estimate. The average common rate and the average private rate are given by
\begin{align}
    \bar{R}_{c,k}&=\mathbb{E}\left[R_{c,k}|\mathbf{\hat{G}}\right] &
    \bar{R}_{k}=\mathbb{E}\left[R_{k}|\mathbf{\hat{G}}\right],
\end{align}
respectively. With the ASR we can obtain the ergodic sum-rate (ESR), which quantifies the performance of the system over a large number of channel realizations and is given by
\begin{equation}
         S_r=\min_{k}\mathbb{E}\left[\bar{R_{c,k}}\right]+\sum_{k=1}^K \mathbb{E}\left[\bar{R}_k\right],\label{system ergodic sum rate}
\end{equation}

\section{Analysis}
\label{analysis}

In this section, we carry out a convexity analysis and a statistical analysis of the proposed algorithms along with an assessment of their computational complexity in terms of floating point operations (FLOPs). Moreover, we derive a bound that establishes that the proposed APA-R algorithm is superior or comparable to the proposed APA algorithm.

\subsection{Convexity analysis}

In this section, we perform a convexity analysis of the optimization problem that gives rise to the proposed APA-R and APA algorithms. In order to establish convexity, we need to compute the second derivative of $\mathbb{E}\left[\varepsilon|\hat{\mathbf{H}}\right]$ with respect to $a_c$ and $a_i$, and then check if it is greater than zero \cite{bertsekas}, i.e.,
\begin{equation}
    \frac{\partial^2 \mathbb{E}\left[\varepsilon|\hat{\mathbf{H}}\right]}{\partial a_c \partial a_c} >0  ~{\rm and} ~ \frac{\partial^2 \mathbb{E}\left[\varepsilon|\hat{\mathbf{H}}\right]}{\partial a_i \partial a_i} >0, ~ i=1,2, \ldots, K
\end{equation}
{Let us first compute $\frac{\partial^2 \mathbb{E}\left[\varepsilon|\hat{\mathbf{H}}\right]}{\partial a_c \partial a_c} $ using the results in \eqref{gradient robust apa ac}:
\begin{equation}
\begin{split}
\frac{\partial}{\partial a_c} \Bigg( \frac{\partial  \mathbb{E}\left[\varepsilon|\hat{\mathbf{H}}\right]}{ \partial a_c} \Bigg)  & = \frac{\partial}{\partial a_c} \Bigg(2\sum_{l=1}^M\sum\limits_{\substack{j=1\\j\neq l}}^M a_c\hat{\phi}^{\left(l,c\right)^*}\hat{\phi}^{\left(j,c\right)}-2\sum_{i=1}^M\Re{\left\{\hat{\phi}^{\left(i,c\right)}\right\}}+4a_c \left(\sum_{i=1}^M\lvert\hat{\phi}^{\left(i,c\right)}\rvert^2+M\sigma_{e,i}^2 \right) \Bigg) \\ & = 2\sum_{l=1}^M\sum\limits_{\substack{j=1\\j\neq l}}^M \hat{\phi}^{\left(l,c\right)^*}\hat{\phi}^{\left(j,c\right)}+4\left(\sum_{i=1}^M\lvert\hat{\phi}^{\left(i,c\right)}\rvert^2+M\sigma_{e,i}^2\lVert\mathbf{p}_c\rVert^2\right). \label{2ndderiv_ac}
\end{split}
\end{equation}
Now let us compute $\frac{\partial^2 \mathbb{E}\left[\varepsilon|\hat{\mathbf{H}}\right]}{\partial a_i \partial a_i} $ using the results in \eqref{gradient robust apa ai}:
\begin{equation}
\begin{split}
\frac{\partial}{\partial a_i} \Bigg( \frac{\partial  \mathbb{E}\left[\varepsilon|\hat{\mathbf{H}}\right]}{ \partial a_i} \Bigg)  & = \frac{\partial}{\partial a_i} \Bigg(4a_i\sum_{l=1}^{M-1}\sum_{q=l+1}^{M} \Re\left[\hat{\phi}^{\left(l,i\right)^*}\hat{\phi}^{\left(q,i\right)}\right]-2 \Re{\left\{\hat{\phi}^{\left(i,i\right)}\right\}}+4a_i\left(\sum_{l=1}^M\lvert\hat{\phi}^{\left(l,i\right)}\rvert^2+M\sigma_{e,i}^2\lVert\mathbf{p}_i\rVert^2\right) \Bigg) \\ & = 4\sum_{l=1}^{M-1}\sum_{q=l+1}^{M} \Re\left[\hat{\phi}^{\left(l,i\right)^*}\hat{\phi}^{\left(q,i\right)}\right]+4\left(\sum_{l=1}^M\lvert\hat{\phi}^{\left(l,i\right)}\rvert^2+M\sigma_{e,i}^2\lVert\mathbf{p}_i\rVert^2\right), \label{2ndderiv_ai}
\end{split}
\end{equation}
Since we have the sum of the strictly convex terms in \eqref{2ndderiv_ac} and \eqref{2ndderiv_ai} the objective function associated with APA-R is strictly convex \cite{bertsekas}. The power constraint is also strictly convex and only scales the powers to be adjusted. In the case of the APA algorithm, the objective function does not employ knowledge of the error variance $\sigma_{e,i}^2$ and remains strictly convex.

\subsection{Bound on the MSE of APA and APA-R}

Let us now show that the proposed APA-R power allocation produces a lower MSE than that of the proposed APA power allocation. The MSE obtained in \eqref{mean square error APA RS} assumes that the transmitter has perfect knowledge of the channel. Under such assumption the optimal coefficients $\mathbf{a}_{o}$ that minimize the error are found. However, under imperfect CSIT the transmitter is unaware of $\tilde{\mathbf{H}}$ and the adaptive algorithm performs the power allocation by employing $\hat{\mathbf{H}}$ instead of $\mathbf{H}$. This results in a power allocation given by $\hat{\mathbf{a}}^{\left(\text{APA}\right)}$ which originates an increase in the MSE obtained. It follows that
\begin{equation}
    \varepsilon\left(\mathbf{H},\mathbf{a}_o\right)\leq\varepsilon\left(\mathbf{H},\hat{\mathbf{a}}^{\left(\text{APA}\right)}\right)
\end{equation}
On the other hand, the robust adaptive algorithm finds the optimal $\mathbf{a}_o^{\left(\text{APA-R}\right)}$ that minimizes $\mathbb{E}\left[\varepsilon\left(\mathbf{H},\mathbf{a}\right)\right|\hat{\mathbf{H}}]$ and therefore takes into account that only partial knowledge of the channel is available. Since the coefficients $\hat{\mathbf{a}}^{\left(\text{APA}\right)}$ and $\mathbf{a}_o^{\left(\text{APA-R}\right)}$ are different, we have
\begin{equation}
    \mathbb{E}\left[\varepsilon\left(\mathbf{H,\mathbf{a}_o^{\left(\text{APA-R}\right)}}\right)|\hat{\mathbf{H}}\right]\leq\mathbb{E}\left[\varepsilon\left(\mathbf{H},\hat{\mathbf{a}}^{\left(\text{APA}\right)}\right)|\hat{\mathbf{H}}\right]
\end{equation}
Note that under perfect CSIT equation \eqref{mean square error APA robust} reduces to \eqref{mean square error APA RS}. In such circumstances $\mathbf{a}_o^{\left(\text{APA}\right)}=\mathbf{a}_o^{\left(\text{APA-R}\right)}$ and therefore both algorithms are equivalent. In the following, we evaluate the performance obtained by the proposed algorithms.
Specifically, we have that $\hat{\mathbf{a}}^{\left(\text{APA}\right)}=\mathbf{a}_o^{\left(\text{APA-R}\right)}+\mathbf{a}_e$, where $\mathbf{a}_e=\left[a_{c,e},a_{1,e},\cdots,a_{M,e}\right]^{\text{T}}$ is the error produced from the assumption that the BS has perfect CSIT.
Then, we have
\begin{align}
    \mathbb{E}\left[\varepsilon^{\left(\text{APA}\right)}-\varepsilon\right.\left.^{\left(\text{APA-R}\right)}\right]&=-2a_{c,e}\sum_{i=1}^M\Re{\left\{\hat{\phi}^{\left(i,c\right)}\right\}}-2\sum_{j=1}^M a_{j,e} \Re{\left\{\hat{\phi}^{\left(j,j\right)}\right\}}+\sum_{l=1}^M\sum\limits_{\substack{j=1\\j\neq l}}^M a_{c,e}^2\hat{\phi}^{\left(l,c\right)^*}\hat{\phi}^{\left(j,c\right)}\nonumber\\
    &+2a_{c,e}^2\left(\sum_{i=1}^M\lvert\hat{\phi}^{\left(i,c\right)}\rvert^2+M\sigma_{e,i}^2\lVert\mathbf{p}_c\rVert^2\right)+2\sum_{i=1}^{M-1}\sum_{q=i+1}^{M}\sum_{r=1}^M a_{r,e}^2\Re\left[\hat{\phi}^{\left(i,r\right)^*}\hat{\phi}^{\left(q,r\right)}\right]\nonumber\\
    &+2\sum_{j=1}^{M}a_{j,e}^2\left(\sum_{l=1}^M\lvert\hat{\phi}^{\left(l,j\right)}\rvert^2+M\sigma_{e,i}^2\lVert\mathbf{p}_j\rVert^2\right).
\end{align}
which is a positive quantity when $-2a_{c,e}\sum_{i=1}^M\Re{\left\{\hat{\phi}^{\left(i,c\right)}\right\}}<2a_{c,e}^2\left(\sum_{i=1}^M\lvert\hat{\phi}^{\left(i,c\right)}\rvert^2+M\sigma_{e,i}^2\lVert\mathbf{p}_c\rVert^2\right)$ and $-2\sum_{j=1}^M a_{j,e} \Re{\left\{\hat{\phi}^{\left(j,j\right)}\right\}}<2\sum_{j=1}^{M}a_{j,e}^2\left(\sum_{l=1}^M\lvert\hat{\phi}^{\left(l,j\right)}\rvert^2+M\sigma_{e,i}^2\lVert\mathbf{p}_j\rVert^2\right)$. The inequalities hold as long as $a_{c,e}\left[\sum_{i=1}^M\left(\Re\left\{\hat\phi^{\left(i,c\right)}\right\}\right)^2+\sum_{i=1}^M\left(\Im\left\{\hat\phi^{\left(i,c\right)}\right\}\right)^2+M\sigma_{e,i}^2\lVert\mathbf{p}_c\rVert^2\right]>\sum_{i=1}^M\Re{\left\{\phi^{\left(i,c\right)}\right\}}$ and $\sum_{j=1}^{M}a_{j,e}\left[\sum_{l=1}^M\left(\Re\left\{\hat\phi^{\left(l,j\right)}\right\}\right)^2+\sum_{l=1}^M\left(\Im\left\{\hat\phi^{\left(l,j\right)}\right\}\right)^2+M\sigma_{e,i}^2\lVert\mathbf{p}_j\rVert^2\right]>\sum_{j=1}^M\Re{\left\{\phi^{\left(j,j\right)}\right\}}$. As the error in the power allocation coefficients grows the left-hand side of the two last inequalities increases. This explains why the proposed APA-R performs better than the proposed APA algorithm.

\subsection{Statistical Analysis}

The performance of adaptive learning algorithms is usually measured in terms of its transient behavior and steady-state behaviour. These measurements provide information about the stability, the convergence rate, and the MSE achieved by the algorithm\cite{Yousef2001,Sayed2003}. 
Let us consider the adaptive power allocation with the update equations given by \eqref{update equation for perfect CSIT}. Expanding the terms of \eqref{update equation for perfect CSIT} for the private streams, we get
\begin{align}
    a_j\left[t+1\right]=&a_j\left[t\right]-4\mu a_j\left[t\right]\sum_{l=1}^{M}\lvert\phi^{\left(l,j\right)}\rvert^2+2\mu\Re\left\{\phi^{\left(j,j\right)}\right\}-4\mu a_j\left[t\right]\sum_{q=1}^{M-1}\sum_{r=q+1}^{M}\Re\left\{\phi^{\left(q,j\right)^*}\phi^{\left(r,j\right)}\right\}.\label{update recursion coeff perfect csit}
\end{align}

Let us define the error between the estimate of the power coefficients and the optimal parameters as follows:
\begin{equation}
    e_{a_j}\left[t\right]=a_j\left[t\right]-a_j^{\left(o\right)},\label{error optimal estimate}
\end{equation}
where $a_j^{\left(o\right)}$ represents the optimal allocation for the $j$th coefficient.

By subtracting \eqref{error optimal estimate} from \eqref{update recursion coeff perfect csit}, we obtain
\begin{align}
    e_{a_j}\left[t+1\right]=&e_{a_j}\left[t\right]-4\mu a_j\left[t\right]\sum_{l=1}^{M}\lvert\phi^{\left(l,j\right)}\rvert^2+2\mu\Re\left\{\phi^{\left(j,j\right)}\right\}-4\mu a_j\left[t\right]\sum_{q=1}^{M-1}\sum_{r=q+1}^{M}\Re\left\{\phi^{\left(q,j\right)^*}\phi^{\left(r,j\right)}\right\}\nonumber\\
    =&e_{a_j}\left[t\right]+2\mu\Re\left\{\phi^{\left(j,j\right)}\right\}-4\mu\left(\sum_{l=1}^{M}\lvert\phi^{\left(l,j\right)}\rvert^2+\sum_{q=1}^{M-1}\sum_{r=q+1}^{M}f_{q,r}^{\left(j\right)}\right)a_j\left[t\right], \label{eq54}
\end{align}
where $f_{q,r}^{\left(j\right)}=\Re\left\{\left(\mathbf{h}_{q,*}\mathbf{p}_j\right)^*\left(\mathbf{h}_{r,*}\mathbf{p}_j\right)\right\}$. Expanding the terms in \eqref{eq54}, we get
\begin{align}
     e_{a_j}\left[t+1\right]=&-4\mu\left(\sum_{l=1}^{M}\lvert\phi^{\left(l,j\right)}\rvert^2+\sum_{q=1}^{M-1}\sum_{r=q+1}^{M}f_{q,r}^{\left(j\right)}\right)a_j^{\left(o\right)}+e_{a_j}\left[t\right]+2\mu\Re\left\{\phi^{\left(j,j\right)}\right\}\nonumber\\
     &-4\mu\left(\sum_{l=1}^{M}\lvert\phi^{\left(l,j\right)}\rvert^2+\sum_{q=1}^{M-1}\sum_{r=q+1}^{M}f_{q,r}^{\left(j\right)}\right)e_{a_j}\left[t\right].\nonumber
\end{align}
Rearranging the terms of the last equation, we obtain
\begin{align}
     e_{a_j}\left[t+ \right.\left. 1\right]
     &=\left\{1-4\mu\left(\sum_{l=1}^{M}\lvert\phi^{\left(l,j\right)}\rvert^2+\sum_{q=1}^{M-1}\sum_{r=q+1}^{M}f_{q,r}^{\left(j\right)}\right)\right\}e_{a_j}\left[t\right]\nonumber\\
     &-4\mu\left(\sum_{l=1}^{M}\lvert\phi^{\left(l,j\right)}\rvert^2+\sum_{q=1}^{M-1}\sum_{r=q+1}^{M}f_{q,r}^{\left(j\right)}\right)a_j^{\left(o\right)}+2\mu\Re\left\{\phi^{\left(j,j\right)}\right\}.\label{eq55}
\end{align}
Equation \eqref{eq55} can be rewritten as follows
\begin{align}
    e_{a_j}\left[t+ \right.\left. 1\right]
    =&\left\{1-4\mu\left(\sum_{l=1}^{M}\lvert\phi^{\left(l,j\right)}\rvert^2+\sum_{q=1}^{M-1}\sum_{r=q+1}^{M}f_{q,r}^{\left(j\right)}\right)\right\}e_{a_j}\left[t\right]\nonumber\\
    &+2\mu\left(\frac{\text{MSE}_{\textrm{min}}\left(a_j^{\left(o\right)}\right)}{a_j^{\left(o\right)}}-a_j^{\left(o\right)}\sum_{q=1}^{M-1}\sum_{r=q+1}^Mf_{q,r}^{\left(j\right)}\right),\label{eq56}
\end{align}
where
\begin{align}
    \text{MSE}_{\rm min}\left(a_j^{\left(o\right)}\right)=2 a_j^{\left(o\right)}\left(a_j^{\left(o\right)}\sum_{i=1}^{M} \lvert\phi^{\left(i,j\right)}\rvert^2-\Re\left\{\phi^{\left(j,j\right)}\right\}\right.\left.+\sum_{q=1}^{M-1}\sum_{r=q+1}^{M}f_{q,r}^{\left(j\right)}\right).
\end{align}

Bu multiplying \eqref{eq56} by $e_a\left[t+1\right]$ and taking the expected value, we obtain
\begin{align}
     \sigma_{e_{a_j}}^2\left[t\right.\left.1+\right]
     =&\left\{1-4\mu\left(\sum_{l=1}^{M}\lvert\phi^{\left(l,j\right)}\rvert^2+\sum_{q=1}^{M-1}\sum_{r=q+1}^{M}f_{q,r}^{\left(j\right)}\right)\right\}^2\sigma_{e_{a_j}}^2\left[t\right]\nonumber\\
    &+4\mu^2\left(\frac{\text{MSE}_{min}\left(a_j^{\left(o\right)}\right)}{a_j^{\left(o\right)}}-a_j^{\left(o\right)}\sum_{q=1}^{M-1}\sum_{r=q+1}^Mf_{q,r}^{\left(j\right)}\right)^2,
\end{align}
where we consider that $\mathbb{E}\left[e_{a_j}\left[i\right]\right]\approx \mathbf{0}$.The previous equation constitutes a geometric series with geometric ratio equal to $1-4\mu\left(\sum_{l=1}^{M}\lvert\phi^{\left(l,j\right)}\rvert^2+\sum_{q=1}^{M-1}\sum_{r=q+1}^{M}f_{q,r}^{\left(j\right)}\right)$. Then, we have
\begin{equation}
    \left\lvert 1-4 \mu\left(\sum_{l=1}^{M}\lvert\phi^{\left(l,j\right)}\rvert^2+\sum_{q=1}^{M-1}\sum_{r=q+1}^{M}f_{q,r}^{\left(j\right)}\right)\right\rvert<1
\end{equation}
Note that from the last equation the step size must fulfill
\begin{equation}
    0<\mu_j<\frac{1}{2\lambda_j},
\end{equation}
with $\lambda_j=\sum_{l=1}^{M}\lvert\phi^{\left(l,j\right)}\rvert^2+\sum_{q=1}^{M-1}\sum_{r=q+1}^{M}f_{q,r}^{\left(j\right)}$.

 For the common power allocation coefficient we have the following recursion:

 \begin{align}
a_c\left[t+1\right]=&a_c\left[t\right]-4\mu a_c\left[t\right]\sum_{j=1}^{M}\lvert\phi^{\left(j,c\right)}\rvert^2-2\mu\sum_{l=1}^{M}\Re\left\{\phi^{\left(l,c\right)}\right\}+2\mu a_c\left[t\right]f^{\left(c\right)},
 \end{align}
where $f^{\left(c\right)}=\sum_{q=1}^{M}\sum_{\substack{r=1\\r\neq q}}^{M}\phi^{\left(q,c\right)^*}\phi^{\left(r,c\right)}$. The error with respect to the optimal power allocation of the common stream is given by
\begin{equation}
    e_c\left[i\right]=a_c{\left[i\right]}-a_c^{\left(o\right)}.
\end{equation}
Following a similar procedure to the one employed for the private streams we arrive at
\begin{align}
    e_c\left[t+1\right]=&\left\{1-4\mu\sum_{j=1}^{M}\lvert\phi^{\left(j,c\right)}\rvert^2-2\mu f^{\left(c\right)}\vphantom{\sum_{j=1}^{M}}\right\}e_c\left[t\right]-2\mu\left(2\sum_{j=1}^{M}\lvert\phi^{\left(j,c\right)}\rvert^2+\mu f^{\left(c\right)}\right)a_c^{\left(o\right)}\nonumber\\
    &-2\mu\sum_{l}^{M}\Re\left\{\phi^{\left(l,c\right)}\right\}.
\end{align}
 Multiplying the previous equation by $e_c\left[t+1\right]$ and taking the expected value leads us to:
 \begin{align}
     \sigma_{e_c}^2\left[t+1\right]=&\left\{1-2\mu\left(2\sum_{j=1}^{M}\lvert\phi^{\left(j,c\right)}\rvert^2-f^{\left(c\right)}\right)\right\}\sigma_{e_c}^2\left[t\right]\nonumber\\
     &-4\left(\frac{\text{MSE}_{\text{min}}\left(a_c^{\left(o\right)}\right)}{a_c^{\left(o\right)}}+a_c^{\left(o\right)}\sum_{j=1}^{M}\Re\left\{\phi^{\left(j,c\right)}\right\}\right)
 \end{align}
 It follows that the geometric ratio of the recursion is equal to $1-2\mu\left(2\sum\limits_{j=1}^{M}\lvert\phi^{\left(j,c\right)}\rvert^2-f^{\left(c\right)}\right)$. Then, the step size must be in the following range:
 \begin{equation}
     0<\mu_c<\frac{1}{\lambda_c},
 \end{equation}
 where
 \begin{equation}
     \lambda_c=2\sum_{j=1}^{M}\lvert\phi^{\left(j,c\right)}\rvert^2-f^{\left(c\right)}
 \end{equation}

 The step-size of the algorithm must be less or equal to $\min\left(\mu_c,\mu_j\right)$ $\forall j \in \left\{1,2,\cdots,M\right\}$. The stability bounds provide useful information on the choice of the step sizes.

Let us now consider the APA-R algorithm. The \textit{a posteriori} error can be expressed as follows:
\begin{align}
    e_{a_j}\left[t+1\right]=&e_{a_j}\left[t\right]+2\mu\Re{\left\{\hat{\phi}^{\left(j,j\right)}\right\}}-4\mu\left(\sum_{l=1}^M\lvert\hat{\phi}^{\left(l,j\right)}\rvert^2+M\sigma_{e,i}^2\lVert\mathbf{p}_j\rVert^2\right.\left.+\sum_{q=1}^{M-1}\sum_{r=q+1}^Mf_{q,r}^{\left(j\right)}\right)a_j\left[t\right].
\end{align}

Expanding the terms of the last equation, we get
\begin{align}
    e_{a_j}\left[t+1\right]=&\left\{1-4\mu\left(\sum_{l=1}^M\lvert\hat{\phi}^{\left(l,j\right)}\rvert^2+M\sigma_{e,i}^2\lVert\mathbf{p}_j\rVert^2\right.\right.\left.\left.+\sum_{q=1}^{M-1}\sum_{r=q+1}^Mf_{q,r}^{\left(j\right)}\right)\right\}e_{a_j}\left[t\right]\nonumber\\
    &-4\mu\left(\sum_{l=1}^M\lvert\hat{\phi}^{\left(l,j\right)}\rvert^2+M\sigma_{e,i}^2\lVert\mathbf{p}_j\rVert^2\right.\left.+\sum_{q=1}^{M-1}\sum_{r=q+1}^Mf_{q,r}^{\left(j\right)}\right)a_j^{\left(o\right)}+2\mu\Re\left\{\hat{\phi}^{\left(j,j\right)}\right\}.
\end{align}
The geometric ratio of the robust APA algorithm is given by $1-4\mu\left(\sum\limits_{l=1}^M\lvert\hat{\phi}^{\left(l,j\right)}\rvert^2+\sigma_{e,i}^2\lVert\mathbf{p}_j\rVert^2\right.$ $\left.+\sum\limits_{q=1}^{M-1}\sum\limits_{r=q+1}^Mf_{q,r}^{\left(j\right)}\right)$. Then, we have that
\begin{equation}
    \left\lvert1-4\mu\left(\sum_{l=1}^{M}\lvert\hat{\phi}^{\left(l,j\right)}\rvert^2+\sigma_{e,i}^2\lVert\mathbf{p}_j\rVert^2+\sum_{q=1}^{M-1}\sum_{r=q+1}^{M}f_{q,r}^{\left(j\right)}\right)\right\rvert<1
\end{equation}
Therefore, the step size must satisfy the following inequality:
\begin{equation}
    0<\mu_j^{\left(\textrm{r}\right)}<\frac{1}{2\lambda^{\left(\textrm{r}\right)}_j},
\end{equation}
where $\lambda^{\left(\textrm{r}\right)}_j=\sum_{l=1}^{M}\lvert\hat{\phi}^{\left(l,j\right)}\rvert^2+\sigma_{e,i}^2\lVert\mathbf{p}_j\rVert^2+\sum_{q=1}^{M-1}\sum_{r=q+1}^{M}f_{q,r}^{\left(j\right)}$. Following a similar procedure for the power coefficient of the common stream lead us to
 \begin{equation}
     0<\mu_c^{\left(\textrm{r}\right)}<\frac{1}{\lambda^{\left(\textrm{r}\right)}_c},
 \end{equation}
 where
  \begin{equation}
     \lambda_c^{\left(\text{r}\right)}=2\sum_{j=1}^{M}\lvert\hat{\phi}^{\left(j,c\right)}\rvert^2+\sigma_{e,i}^2\lVert\mathbf{p}_j\rVert^2-f^{\left(c\right)}
 \end{equation}
As in the previous case, the step size is chosen using $\min\left(\mu^{\left(\textrm{r}\right)}_c,\mu^{\left(\textrm{r}\right)}_j\right)$ $\forall j \in \left\{1,2,\cdots,M\right\}$. The variable $\lambda_c^{\left(\textrm{r}\right)}$ has an additional term given by $\sigma_{e,i}^2\lVert\mathbf{p}_j\rVert^2$ when compared to $\lambda_c$ of the APA algorithm. In this sense, the upper bound of (70) is smaller than the bound in (65). In other words, the step size of APA-R takes smaller values than the step size of APA.

\subsection{Computational Complexity}
In this section the number of FLOPs performed by the proposed algorithms is computed. For this purpose let us consider the following results to simplify the analysis. Consider the complex vector $\mathbf{z}_1$ and $\mathbf{z}_2 \in \mathbb{C}^{n}$. Then, we have the following results:
\begin{itemize}
    \item The product $\mathbf{z}_1^{\text{T}}\mathbf{z}_2$ requires $8n-2$ FLOPs.
    \item The term $\lVert \mathbf{z}_1\rVert^2$ requires $7n-2$ FLOPs
\end{itemize}

The gradient in equation \eqref{gradient RS perfect CSIT private streams} requires the computation of three terms. The first term, which is given by $4 a_c \sum\limits_{i=1}^M\lvert\mathbf{h}_{i,*}\mathbf{p}_c\rvert^2$ needs a total of $8 N_tM +2M+1$ FLOPs. The evaluation of the second term results in $8N_tM+M$. For the last term we have a total of $\left(9M^2-9M+2\right)/2$. Considering a system where $N_t=N_r=M=n$ we have that the number of FLOPs required by the proposed adaptive algorithm is given by $\frac{41}{2}n^3+19n^2+\frac{5}{2}n+4$.

The computational complexity of the gradients in \eqref{gradient robust apa ac} and \eqref{gradient robust apa ai} can be computed in a similar manner. However, in this case we have an additional term given by $4 a_c M\sigma_{e,i}^2\lVert \mathbf{p}_i\rVert^2$, which requires a total of $7N_t+2$ FLOPs. Then, the robust adaptive algorithm requires a total of $\frac{41}{2}n^3+19n^2+\frac{19}{2}n+6$. It is important to mention that the computational complexity presented above represents the number of FLOPs of the adaptive algorithm per iteration.

In contrast, the optimal power allocation for the conventional SDMA system requires the exhaustive search over $\mathbf{A}$ with a fine grid. Given a system with $12$ streams and a grid step of $0.001$, the exhaustive search would require the evaluation of $5005000$ different power allocation matrices for each channel realization. In contrast, the adaptive approaches presented require only the computation of around $30$ iterations. Furthermore, the complexity of  the exhaustive search for an RS system is even higher since the search is perform over $\mathbf{A}^{\left(\text{RS}\right)}$, which additionally contains the power allocated to the common stream.

Table \ref{Computational complexity power allocation} summarizes the computational complexity of the proposed algorithms employing the big $\mathcal{O}$ notation. In  Table \ref{Computational complexity power allocation}, $I_o$ denotes the number of points of the grid given a step size, $I_w$ refers to the number of iterations of the alternating procedure and $I_a$ denotes the number of iterations for the adaptive algorithms. It is worth noting that $I_o>>I_a$. Moreover, the inner iterations employed in the WMMSE approach are much more demanding than the iterations of the proposed APA and APA-R algorithms. Fig \ref{Complexity} shows the computational complexity in terms of FLOPS assuming that the number of transmit antennas increases. The term CF represents the closed-form power allocation in \cite{Dai2016a}, which requires the inversion of an $N_t\times N_t$ matrix. The step of the grid was set to $0.01$ for the ES and the number of iterations to $30$ for the WMMSE, APA and APA-R approaches. Note that in practice the WMMSE iterates continuously until meeting a predefined accuracy. In general, it requires more than $30$ iterations. It is also important to point out that the cost per iteration of the adaptive approaches can be reduced after the first iteration. The precoders and the channel are fixed given a transmission block. Therefore, after the first iteration, we can avoid the multiplication of the precoder by the channel matrix in the update equation. In contrast, the WMMSE must perform the whole procedure again since the precoders are updated between iterations. This is illustrated in Fig. \ref{ComplexityPerIteration}.

\begin{table}[H]
\caption{Computational complexity of the power allocation algorithms.}
\begin{center}
\vspace{-.3cm}
\begin{tabular}{ p{4 cm} c}
\hline
\hline
 Technique & $\mathcal{O}$\\
\hline
\rule{0pt}{3ex}
 SDMA-ES & $\mathcal{O}\left(N_t I_o^2 M^3\right)$\\
 \rule{0pt}{3ex}
 WMMSE & $\mathcal{O}\left(I_w N_t M^3\right)$\\
\rule{0pt}{3ex}
 RS-ES & $\mathcal{O}\left(N_t I_o^2 (M+1)^3\right)$\\
\rule{0pt}{3ex}
 RS-APA & $\mathcal{O}\left(I_a N_t (M+1)^2\right)$\\
\rule{0pt}{3ex}
 RS-APA-R & $\mathcal{O}\left(I_a N_t (M+1)^2\right)$\\
 \rule{0pt}{3ex}
 CF\cite{Dai2016a} & $\mathcal{O}\left( N_t^3\right)$\\
\hline\label{Computational complexity power allocation}
\end{tabular}
\end{center}
\end{table}
\vspace{-2em}

\begin{figure}[h]
\centering
    \begin{subfigure}[b]{0.45\textwidth}
    \centering
    \includegraphics[height=5.5cm, width=0.98\columnwidth]{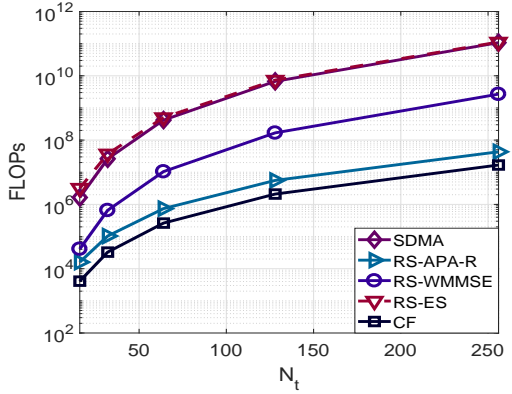}
    \caption{Number of FLOPs required by different power allocation algorithms considering a MU-MIMO system with an increasing number of antennas.}
    \label{Complexity}
    \end{subfigure}
    \quad
    \begin{subfigure}[b]{0.45\textwidth}.
    \includegraphics[height=5.6cm, width=0.98\columnwidth]{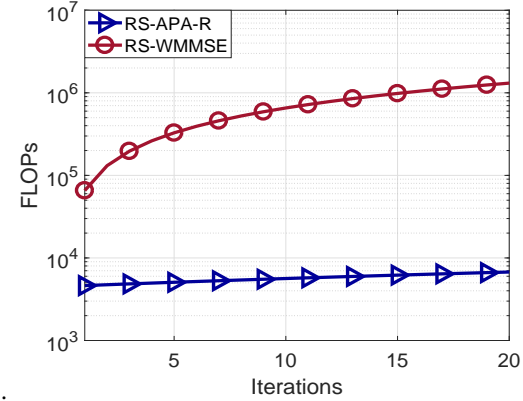}
    \caption{Number of FLOPs required per iteration considering a MU-MIMO system with $Nt=16$ and $M=16$.}
    \label{ComplexityPerIteration}
    \end{subfigure}
    \caption{Computational Complexity}

\end{figure}


\section{Simulations}\label{c5 section simulations}

In this section, the performance of the proposed APA-R and APA algorithms is assessed against existing power allocation approaches, namely, the ES, the WMMSE \cite{JoudehClerckx2016}, the closed-form approach of \cite{Dai2016a}, the power allocation obtained directly from the precoders definition which is denoted here as random power allocation, and the uniform power allocation (UPA) approaches. Unless otherwise specified, we consider an RS MU-MIMO system where the BS is equipped with four antennas and transmits data to two users, each one equipped with two antennas. The inputs are statistically independent and follow a Gaussian distribution. A flat fading Rayleigh channel, which remains fixed during the transmission of a packet, is considered, we assume additive white Gaussian noise with zero mean and unit variance, and the SNR varies with $E_{tr}$. For all the experiments, the common precoder is computed by employing a SVD over the channel matrix, i.e. $\hat{\mathbf{H}}=\mathbf{S}\mathbf{\Psi}\mathbf{V}^H$. Then we set the common precoder equal to the first column of matrix $\mathbf{V}$, i.e., $\mathbf{p}_c=\mathbf{v}_1$.

In the first example, we consider the transmission under imperfect CSIT. The proposed APA-R algorithm is compared against the closed form expression from \cite{Dai2016a} and against ES with random and UPA algorithms. The first ES approach fixes a random power allocation for the private streams and then an exhaustive search is carried out to find the best power allocation for the common stream. The second scheme considers that the power is uniformly distributed among private streams and then performs an exhaustive search to find the optimum value for $a_c$. Fig. \ref{C5 Figure3} shows the performance obtained with a MF. Although ES obtains the best performance, it requires a great amount of resources in terms of computational complexity. Moreover, we can see that the closed-form power allocation does not allocate power to the common message in the low SNR regime. The reason for this behavior is that this power allocation scheme was derived for massive MIMO environments where the excess of transmit antennas gets rid of the residual interference at low SNR and no common message is required. As the SNR increases the residual interference becomes larger and the algorithm allocates power to the common stream.

\begin{figure}[h]
\begin{center}
\includegraphics[scale=0.55]{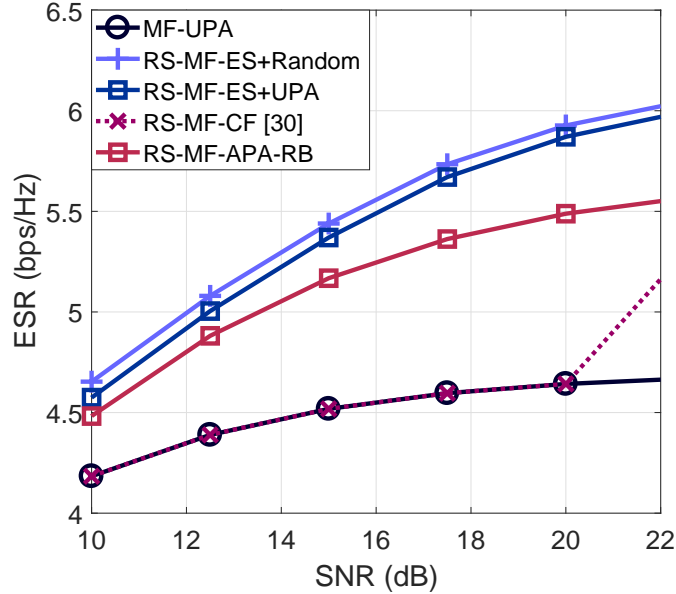}
\vspace{-1.0em}
\caption{Sum-rate performance of RS-MF precoding scheme, $N_t=4$, $N_k=4$, $K=1$, and $\sigma_e^2=0.05$.}
\label{C5 Figure3}
\end{center}
\end{figure}

In the next example,  the ZF precoder has been considered. The results illustrated in Fig. \ref{C5 Figure4} show that the techniques that perform ES, which are termed as RS-ZF-ES+Random and RS-ZF-ES-UPA, have the best performance. However, the APA and APA-R adaptive algorithms obtain a consistent gain when compared to the conventional ZF algorithm.

\begin{figure}[h]
\begin{center}
\includegraphics[scale=0.45]{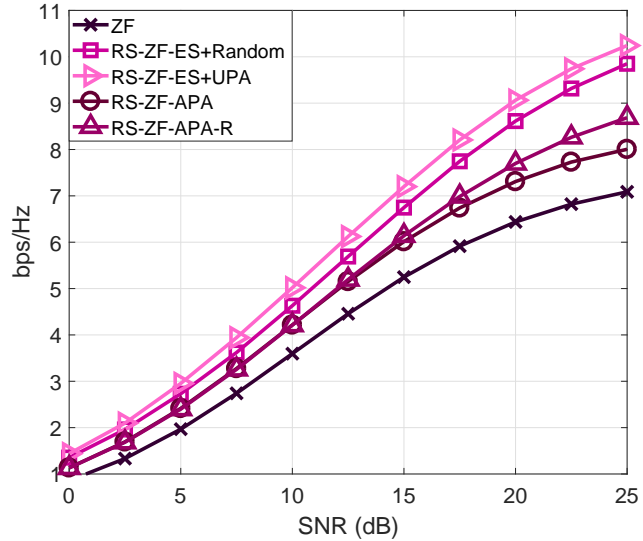}
\vspace{-1.0em}
\caption{Sum-rate performance of RS-ZF precoding scheme, $N_t=4$, $N_k=2$, $K=2$, and $\sigma_e^2=0.1$.}
\label{C5 Figure4}
\end{center}
\end{figure}

In Fig. \ref{C5_7_D} we employed the MMSE precoder. In this case, an exhaustive search was performed to obtain the best power allocation coefficients for all streams. The technique from [27] was also considered and is denoted in the figure as RS-WMMSE. We can observe that the best performance is obtained by ES. The proposed APA and APA-R algorithms obtain a consistent gain when compared to the conventional MMSE precoder. The performance is worse than that of ES and the RS-WMMSE, but the computational complexity is also much lower.

\begin{figure}[h]
\begin{center}
\includegraphics[scale=0.45]{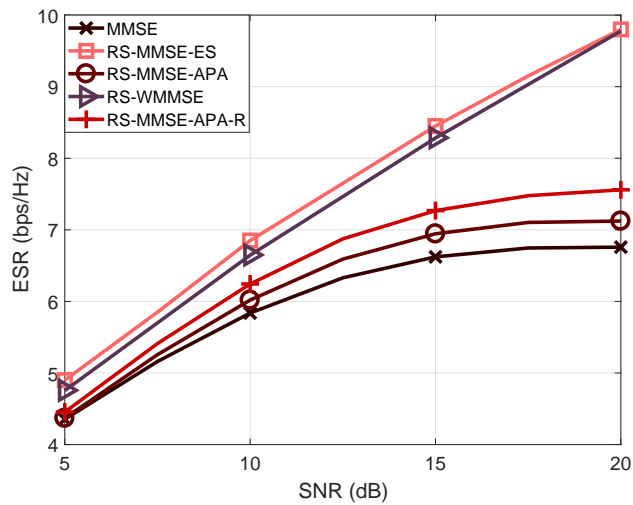}
\vspace{-1.0em}
\caption{Sum-rate performance of RS-MMSE precoding scheme, $N_t=4$, $N_k=2$, $K=2$, and $\sigma_e^2=0.2$.}
\label{C5_7_D}
\end{center}
\end{figure}

In the next example, we evaluate the performance of the proposed APA and APA-R techniques as the error in the channel estimate becomes larger. For this scenario, we consider a fixed SNR equal to 20 dB. Fig. \ref{VarErr} depicts the results of different power allocation techniques. The results show that the APA-R performs better than the APA as the variance of the error increases.

\begin{figure}[h]
\begin{center}
\includegraphics[scale=0.45]{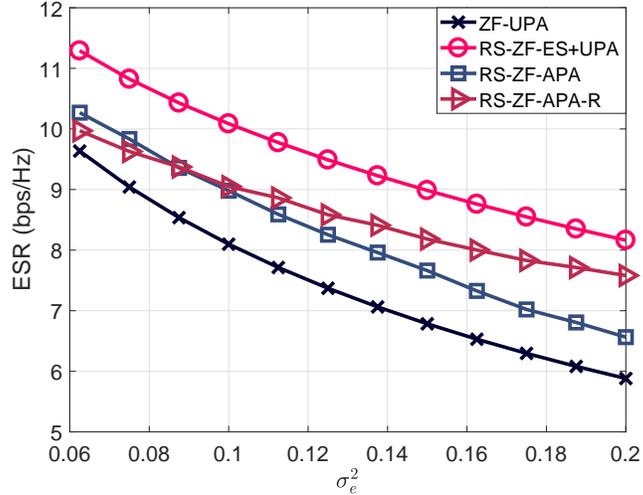}
\vspace{-1.0em}
\caption{Sum-rate performance of RS-ZF precoding scheme, $N_t=4$, $N_k=2$, and $K=2$.}
\label{VarErr}
\end{center}
\end{figure}

Let us now consider the ESR obtained versus the number of iterations, which is shown in Fig. \ref{MSEperIteration}. The step size of the adaptive algorithms was set to $0.004$ and the SNR to $10$ dB. Fig. \ref{MSEperIteration} shows that APA and APA-R obtain better performance than WMMSE with few iterations, i.e., with reduced cost. Recall that the cost per iteration is much lower for the adaptive algorithms.

\begin{figure}[h]
\begin{center}
\includegraphics[scale=0.55]{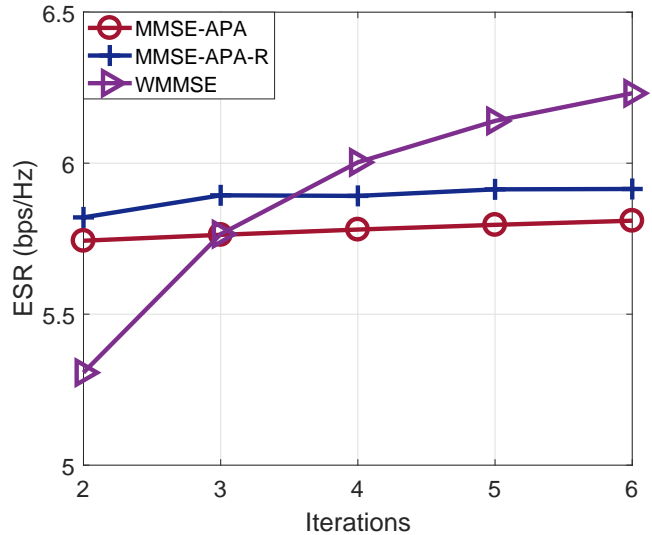}
\vspace{-1.0em}
\caption{Sum-rate performance of RS-ZF precoding scheme, $N_t=4$, $N_k=1$,$K=4$, and $\sigma_e^2=0.2$.}
\label{MSEperIteration}
\end{center}
\end{figure}

In Fig. \ref{acPow} we can notice the power allocated to the common stream. For this simulation we consider the same setup as in the previous simulation. We can observe that the parameter $a_c$ increases with the SNR. In other words, as the MUI becomes more significant, more power is allocated to the common stream. We can also notice that the proposed APA-R algorithm allocates more power to the common stream than that of the APA algorithm.

\begin{figure}[h]
\begin{center}
\includegraphics[scale=0.45]{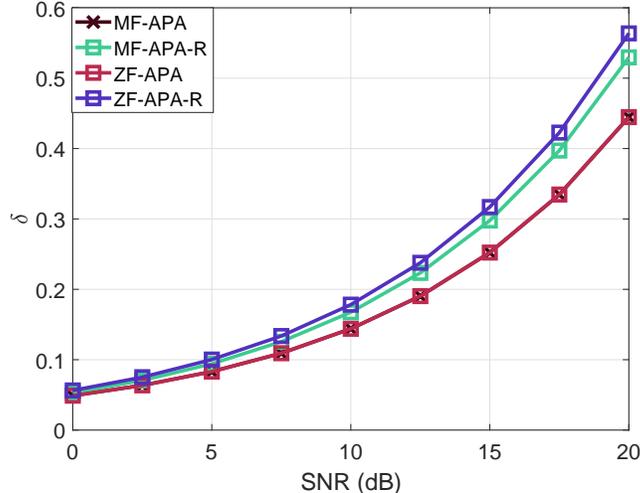}
\vspace{-1.5em}
\caption{Power allocated to the common stream, $N_t=4$, $N_k=2$, $K=2$.}
\label{acPow}
\vspace{-1.5em}
\end{center}
\end{figure}

In the last example, we consider the ZF precoder in a MU-MIMO system where the BS is equipped with $24$ transmit antennas. The information is sent to $24$ users which are randomly distributed over the area of interest. Fig. \ref{C5 Figure6} shows the results obtained by employing the proposed APA and APA-R algorithms. Specifically, it can be noticed that the RS system equipped with APA-R can obtain a gain of up to $20 \%$ over that random allocation and up to $50 \%$ over that of a conventional MU-MIMO system with random allocation.

\begin{figure}[h]
\begin{center}
\includegraphics[scale=0.5]{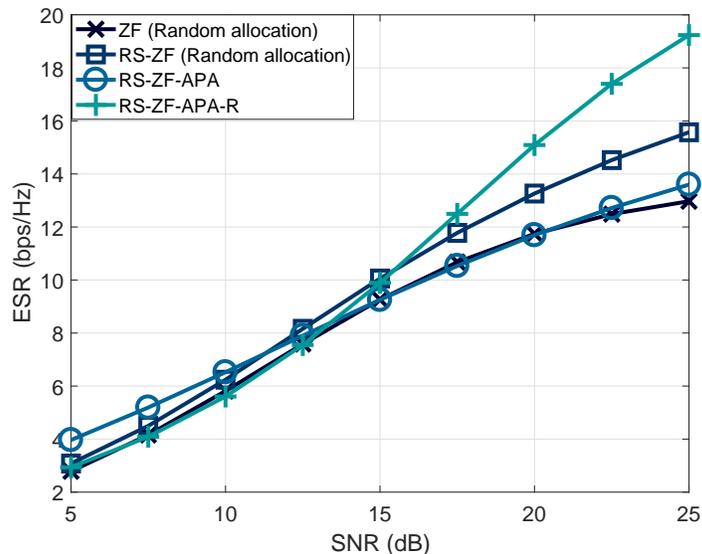}
\vspace{-1.5em}
\caption{Sum-rate performance of RS-ZF precoding scheme, $N_t=24$, $N_k=1$, $K=24$, and $\sigma_e^2=0.1$.}
\label{C5 Figure6}
\end{center}
\end{figure}

\vspace{-2.15em}
\section{Conclusion}
In this work, adaptive power allocation techniques have been developed for RS-MIMO and conventional MU-MIMO systems. Differently to optimal and WMMSE power allocation often employed for RS-based systems that are computationally very costly, the proposed APA and APA-R algorithms have low computational complexity and require fewer iterations for new transmission blocks, being suitable for practical systems. Numerical results have shown that the proposed power allocation algorithms, namely APA and APA-R, are not very far from the performance of exhaustive search with uniform power allocation. Furthermore, the proposed robust technique, i.e., APA-R, increases the robustness of the system against CSIT imperfections.


%
\appendices
\section{Proof of the MSE for the APA-R}\label{Appendix MSE APA-R}
In what follows the derivation of the MSE for the APA-R algorithm is detailed. Let us first expand the MSE, which is given by
\begin{align}
    \mathbb{E}\left[\varepsilon\lvert\hat{\mathbf{H}}\right]=&\mathbb{E}\left[\left(\mathbf{s}^{\left(\text{RS}\right)}-\mathbf{y}'\right)^H\left(\mathbf{s}^{\left(\text{RS}\right)}-\mathbf{y}'\right)\lvert\hat{\mathbf{H}}\right]\nonumber\\
    =&\underbrace{\mathbb{E}\left[\mathbf{s}^{\left(\text{RS}\right)^H}\mathbf{s}^{\left(\text{RS}\right)}\lvert\hat{\mathbf{H}}\right]}_{T_1}-\underbrace{\mathbb{E}\left[\mathbf{s}^{\left(\text{RS}\right)^H}\mathbf{y}'\lvert\hat{\mathbf{H}}\right]}_{T_2}-\underbrace{\mathbb{E}\left[\mathbf{y'}^H\mathbf{s}^{\left(\text{RS}\right)}\lvert\hat{\mathbf{H}}\right]}_{T_3}+\underbrace{\mathbb{E}\left[\mathbf{y'}^H\mathbf{y}'\lvert\hat{\mathbf{H}}\right]}_{T_4}.\label{mean square error terms for RS APA Robust}
\end{align}

The first term of \eqref{mean square error terms for RS APA Robust} is independent from $\hat{\mathbf{H}}$ and can be reduced to the following expression:
\vspace{-2em}
\begin{align}
    \mathbb{E}\left[\mathbf{s}^{\left(\text{RS}\right)^H}\mathbf{s}^{\left(\text{RS}\right)}\right]=&\mathbb{E}\left[s_c^*s_c\right]+\mathbb{E}\left[s_1^*s_1\right]\cdots+\mathbb{E}\left[s_M^*s_M\right],=M+1.\label{c5 term1}
\end{align}
The second term given by $T_2$ requires the computation of the following term:
\begin{align}
    \mathbf{s}^{\left(\text{RS}\right)^H}\mathbf{y}'=& s_c^* y_c+s_1^*y_1+\cdots+s_M^*y_{M},\nonumber\\
    =&s_c^*\sum_{l=1}^{M}\left[a_c s_c\left(\hat{\mathbf{h}}_{l,*}+\tilde{\mathbf{h}}_{l,*}\right)\mathbf{p}_c+\sum_{j=1}^{M}a_j s_j \left(\hat{\mathbf{h}}_{l,*}+\tilde{\mathbf{h}}_{l,*}\right)\mathbf{p}_j+ n_l\right]\nonumber\\
    &+s_1^*\left[a_c s_c \left(\hat{\mathbf{h}}_{1,*}+\tilde{\mathbf{h}}_{1,*}\right)\mathbf{p}_c+\sum_{l=1}^{M}a_l s_l\left(\hat{\mathbf{h}}_{1,*}+\tilde{\mathbf{h}}_{1,*}\right)\mathbf{p}_l+n_1\right]\nonumber\\
    &+\cdots+s_M^*\left[a_c s_c \left(\hat{\mathbf{h}}_{M,*}+\tilde{\mathbf{h}}_{M,*}\right)\mathbf{p}_c+\sum_{l=1}^{M}a_l s_l\left(\hat{\mathbf{h}}_{M,*}+\tilde{\mathbf{h}}_{M,*}\right)\mathbf{p}_l+n_M\right].\label{c5 term2 new}
\end{align}
By evaluating the expected value of the different terms in \eqref{c5 term2 new} we get the following quantities:
\begin{align}
     \mathbb{E}\left[s_i^*y_i|\hat{\mathbf{H}}\right]&=\mathbb{E}\left[s_i^*\left\{a_c s_c \left(\hat{\mathbf{h}}_{i,*}+\tilde{\mathbf{h}}_{i,*}\right)\mathbf{p}_c+\sum_{l=1}^{M}a_l s_l\left(\hat{\mathbf{h}}_{i,*}+\tilde{\mathbf{h}}_{i,*}\right)\mathbf{p}_l+n_i\right\}\lvert\hat{\mathbf{H}}\right],\nonumber\\
    &=a_i\hat{\phi}^{\left(i,i\right)}.\\
    \mathbb{E}\left[s_c^*y_c|\hat{\mathbf{H}}\right]&=\mathbb{E}\left[s_c^*\sum_{l=1}^{M}\left\{a_c s_c\left(\hat{\mathbf{h}}_{l,*}+\tilde{\mathbf{h}}_{l,*}\right)\mathbf{p}_c+\sum_{j=1}^{M}a_j s_j \left(\hat{\mathbf{h}}_{l,*}+\tilde{\mathbf{h}}_{l,*}\right)\mathbf{p}_j+ n_l\right\}\lvert\hat{\mathbf{H}}\right],\nonumber\\
      &=a_c\sum_{i=1}^M \hat{\phi}^{\left(i,c\right)},
\end{align}where $\hat{\phi}^{\left(i,q\right)}=\hat{\mathbf{h}}_{i,*}\mathbf{p}_q$ and $\hat{\phi}^{\left(i,c\right)}=\hat{\mathbf{h}}_{i,*}\mathbf{p}_c$. These expressions allow us to compute $T_2$, which is expressed by
\begin{equation}
    \mathbb{E}\left[\mathbf{s}^{\left(\text{RS}\right)^H}\mathbf{y}'|\hat{\mathbf{H}}\right]=a_c\sum_{i=1}^M\hat{\phi}^{\left(i,c\right)}+\sum_{j=1}^{M}a_j\hat{\phi}^{\left(j,j\right)}.\label{c5 term 2 robust}
\end{equation}
The third term can be calculated in a similar manner and is given by
\begin{align}
    \mathbb{E}\left[\mathbf{y'}^H\mathbf{s}^{\left(\text{RS}\right)}\right.&\left.|\hat{\mathbf{H}}\right]=a_c\sum_{i=1}^M\hat{\phi}^{\left(i,c\right)^*}+\sum_{j=1}^{M}a_j\hat{\phi}^{\left(j,j\right)^*}.\label{c5 term 3 robust}
\end{align}

The last term of equation \eqref{mean square error terms for RS APA Robust} requires the computation of several quantities. Let us first consider the following quantity:
\begin{align}
    \mathbf{y'}^H\mathbf{y}'=&y_c^* y_c+y_1^*y_1+\cdots+y_M^*y_M=\left(\sum_{l=1}^M y_l^*\right)\left(\sum_{j=1}^M y_j\right)+\sum_{i=1}^M y_i^*y_i.
\end{align}
Taking the expected value on the last equation results in
\begin{align}
    \mathbb{E}\left[ \mathbf{y'}^H\mathbf{y}'|\hat{\mathbf{H}}\right]=&\sum_{l=1}^M\sum_{j=1}^M\mathbb{E}\left[y_l^* y_j|\hat{\mathbf{H}}\right]+\sum_{i=1}^{M}\mathbb{E}\left[y_i^* y_i|\hat{\mathbf{H}}\right],\nonumber\\
    =&\sum_{l=1}^M\sum\limits_{\substack{j=1\\j\neq l}}^M\mathbb{E}\left[y_l^* y_j|\hat{\mathbf{H}}\right]+2\sum_{i=1}^{M}\mathbb{E}\left[y_i^* y_i|\hat{\mathbf{H}}\right],\label{c5 term4 complete}
\end{align}

\begin{align}
    \mathbb{E}\left[y_i^* y_i\lvert\hat{\mathbf{H}}\right]=&\mathbb{E}\left[\left\lvert a_c s_c \mathbf{h}_{i,*}\mathbf{p}_c+\sum_{q=1}^{M}a_q s_q \mathbf{h}_{i,*}\mathbf{p}_q+ n_i\right\rvert^2\rvert\hat{\mathbf{H}}\right],\nonumber\\
    =&\mathbb{E}\left[a_c^2\lvert s_c\rvert^2\left\lvert\hat{\phi}^{\left(i,c\right)}+\tilde{\phi}^{\left(i,c\right)}\right\rvert^2+\sum_{q=1}^{M}a_q\lvert s_q\rvert^2\left\lvert\hat{\phi}^{\left(i,q\right)}+\tilde{\phi}^{\left(i,q\right)}\right\rvert^2+\lvert n_i\rvert^2 \lvert \hat{\mathbf{H}}\right],
\end{align}
where  $\tilde{\phi}^{\left(i,c\right)}=\tilde{\mathbf{h}}_{i,*}\mathbf{p}_c$ and $\tilde{\phi}^{\left(i,q\right)}=\tilde{\mathbf{h}}_{i,*}\mathbf{p}_q$. Expanding the terms of the last equation results in
\begin{align}
    \mathbb{E}\left[y_i^* y_i\lvert\hat{\mathbf{H}}\right]=&\mathbb{E}\left[\sum_{q=1}^M a_q^2\lvert s_q\rvert^2\left(\lvert\hat{\phi}^{\left(i,q\right)}\rvert^2+2\Re\left[\hat{\phi}^{\left(i,q\right)^*}\tilde{\phi}^{\left(i,q\right)}\right]+\lvert\tilde{\phi}^{\left(i,q\right)}\rvert^2\right)\lvert\hat{\mathbf{H}}\right]\nonumber\\
    &+\mathbb{E}\left[a_c^2\lvert s_c\rvert^2\left(\lvert\hat{\phi}^{\left(i,c\right)}\rvert^2+2\Re\left[\hat{\phi}^{\left(i,c\right)^*}\tilde{\phi}^{\left(i,c\right)}\right]+\lvert\tilde{\phi}^{\left(i,c\right)}\rvert^2\right)|\hat{\mathbf{H}}\right]+\sigma_n^2.
\end{align}
Since the entries of $\tilde{\mathbf{h}}_{i,*}~~\forall i$ are uncorrelated with zero mean and also independent from $\mathbf{s}^{\left(\text{RS}\right)}$, we have
\begin{align}
    \mathbb{E}\left[y_i^* y_i|\hat{\mathbf{H}}\right]=&a_c^2\left(\lvert\hat{\phi}^{\left(i,c\right)}\rvert^2+\mathbb{E}\left[\lvert\tilde{\phi}^{\left(i,c\right)}\rvert^2\lvert\hat{\mathbf{H}}\right]\right)+\sum_{q=1}^M a_q^2\left(\lvert\hat{\phi}^{\left(i,q\right)}\rvert^2+\mathbb{E}\left[\lvert\tilde{\phi}^{\left(i,q\right)}\rvert^2\lvert\hat{\mathbf{H}}\right]\right)+\sigma_n^2.\label{c5 term4 robust part1}
\end{align}
Note that $\lvert\tilde{\phi}^{\left(i,c\right)}\rvert^2$ and $\lvert\tilde{\phi}^{\left(i,q\right)}\rvert^2$ are independent from $\hat{\mathbf{H}}$. Thus, we get
\begin{align}
    \mathbb{E}\left[\lvert\tilde{\phi}^{\left(i,c\right)}\rvert^2\right]=&\lvert p^{\left(c\right)}_1\rvert^2\mathbb{E}\left[\tilde{h}_{i,1}^*\tilde{h}_{i,1}\right]+\lvert p^{\left(c\right)}_2\rvert^2\mathbb{E}\left[\tilde{h}_{i,2}^*\tilde{h}_{i,2}\right]+\cdots+\lvert p^{\left(c\right)}_{N_t}\rvert^2\mathbb{E}\left[\tilde{h}_{i,N_t}^*\tilde{h}_{i,N_t}\right],\nonumber\\
    =&\lvert p^{\left(c\right)}_1\rvert^2\sigma_{e,i}^2+\lvert p^{\left(c\right)}_2\rvert^2\sigma_{e,i}^2+\cdots\lvert p^{\left(c\right)}_{N_t}\rvert^2\sigma_{e,i}^2,\nonumber\\
    =&\sigma_{e,i}^2\lVert\mathbf{p}_c\rVert^2,
\end{align}
and similarly $\mathbb{E}\left[\lvert\tilde{\phi}^{\left(i,q\right)}\rvert^2\right]=\sigma_{e,i}^2\lVert\mathbf{p}_q\rVert^2.$
Then, \eqref{c5 term4 robust part1} turns into
\begin{align}
    \mathbb{E}\left[ y_i^* y_i|\hat{\mathbf{H}}\right]=&a_c^2\left(\lvert\hat{\phi}^{\left(i,c\right)}\rvert^2+\sigma_{e,i}^2\lVert\mathbf{p}_c\rVert^2\right)+\sum_{j=1}^{K}a_j^2\left(\lvert\hat{\phi}^{\left(i,j\right)}\rvert^2+\sigma_{e,i}^2\lVert\mathbf{p}_j\rVert^2\right)+\sigma_n^2.\label{c5 corr received signal same ant robust}
\end{align}

Let us now evaluate the expected value of $y_l^*y_j$ when $l\neq j$, which results in
\begin{align}
    \mathbb{E}\left[y_l^* y_j\lvert\hat{\mathbf{H}}\right]=&\mathbb{E}\left[\left(a_c s_c \mathbf{h}_{l,*}\mathbf{p}_c+\sum_{q=1}^{M}a_q s_q \mathbf{h}_{l,*}\mathbf{p}_q+ n_l\right)^*\right.\times\left.\left(a_c s_c \mathbf{h}_{j,*}\mathbf{p}_c+\sum_{r=1}^{M}a_r s_r \mathbf{h}_{j,*}\mathbf{p}_r+ n_j\right)\lvert\hat{\mathbf{H}}\right],\nonumber\\
     =&\sum_{q=1}^M a_q^2\mathbb{E}\left[\hat{\phi}^{\left(l,q\right)^*}\hat{\phi}^{\left(j,q\right)}+ \hat{\phi}^{\left(l,q\right)^*}\tilde{\phi}^{\left(j,q\right)}+\tilde{\phi}^{\left(l,q\right)^*}\hat{\phi}^{\left(j,q\right)}+\tilde{\phi}^{\left(l,q\right)^*}\tilde{\phi}^{\left(j,q\right)}\lvert\hat{\mathbf{H}}\right].\nonumber\\
    &+a_c^2\mathbb{E}\left[\hat{\phi}^{\left(l,c\right)^*}\hat{\phi}^{\left(j,c\right)}+\hat{\phi}^{\left(l,c\right)^*}\tilde{\phi}^{\left(j,c\right)}+\tilde{\phi}^{\left(l,c\right)^*}\hat{\phi}^{\left(j,c\right)}+\tilde{\phi}^{\left(l,c\right)^*}\tilde{\phi}^{\left(j,c\right)}\lvert\hat{\mathbf{H}}\right].
\end{align}

Remark that $\tilde{\mathbf{h}}_l$ and $\tilde{\mathbf{h}}_j$ are independent $\forall~~l\neq j$ with zero mean. Thus, the last equation is reduced to
\begin{align}
   \mathbb{E}\left[y_l^* y_j|\hat{\mathbf{H}}\right]=&a_c^2\hat{\phi}^{\left(l,c\right)^*}\hat{\phi}^{\left(j,c\right)}+\sum_{q=1}^M a_q^2\hat{\phi}^{\left(l,q\right)^*}\hat{\phi}^{\left(j,q\right)}.\label{c5 corr signal from different ant robust}
\end{align}

Equations \eqref{c5 corr received signal same ant robust} allow us to compute the second term of equation \eqref{c5 term4 complete}, which is given by
\begin{align}
    \sum_{i=1}^M\mathbb{E}\left[y_i^*y_i|\right.\left.\hat{\mathbf{H}}\right]
    =&\sum_{j=1}^{M}a_j^2\left(\sum_{l=1}^M\lvert\hat{\phi}^{\left(l,j\right)}\rvert^2+M\sigma_{e_i}^2\lVert\mathbf{p}_j\rVert^2\right)+a_c^2\left(\sum_{i=1}^M\lvert\hat{\phi}^{\left(i,c\right)}\rvert^2+M\sigma_{e,i}^2\lVert\mathbf{p}_c\rVert^2\right)+M\sigma_n^2.\label{c5 term 4 part 1 robust}
    \end{align}
On the other hand, \eqref{c5 corr signal from different ant robust} allow us to obtain the first term of \eqref{c5 term4 complete}, which results in
\begin{equation}
    \sum_{l=1}^M\sum\limits_{\substack{j=1\\j\neq l}}^M\mathbb{E}\left[y_l^* y_j|\hat{\mathbf{H}}\right]=\sum_{l=1}^M\sum\limits_{\substack{j=1\\j\neq l}}^M\left(a_c^2\hat{\phi}^{\left(l,c\right)^*}\hat{\phi}^{\left(j,c\right)}+\sum_{q=1}^M a_q^2\hat{\phi}^{\left(l,q\right)^*}\hat{\phi}^{\left(j,q\right)}\right)
\end{equation}
Applying the property $\hat{\phi}^{\left(l,q\right)^*}\hat{\phi}^{\left(j,q\right)}+\hat{\phi}^{\left(l,q\right)}\hat{\phi}^{\left(j,q\right)^*}=2\Re\left[\hat{\phi}^{\left(l,q\right)^*}\hat{\phi}^{\left(j,q\right)}\right]$, we can simplify half of the sums from the triple summation, i.e.,
\begin{equation}
\sum_{l=1}^M\sum\limits_{\substack{j=1\\j\neq l}}^M\sum_{q=1}^Ma_q^2\hat{\phi}^{\left(l,q\right)^*}\hat{\phi}^{\left(j,q\right)}=2\sum_{l=1}^{M-1}\sum_{j=i+1}^{M}\sum_{q=1}^M a_q^2\Re\left[\hat{\phi}^{\left(l,q\right)^*}\hat{\phi}^{\left(j,q\right)}\right].\label{c5 triple summation explained new}
\end{equation}
It follows that
\begin{align}
    \sum_{l=1}^M\sum\limits_{\substack{j=1\\j\neq l}}^M\mathbb{E}\left[y_l^* y_j|\hat{\mathbf{H}}\right]=&2\sum_{l=1}^{M-1}\sum_{j=i+1}^{M}\sum_{q=1}^M a_r^2\Re\left[\hat{\phi}^{\left(l,q\right)^*}\hat{\phi}^{\left(j,q\right)}\right]+\sum_{l=1}^M\sum\limits_{\substack{j=1\\j\neq l}}^M a_c^2\hat{\phi}^{\left(l,c\right)}\hat{\phi}^{\left(j,c\right)}.\label{c5 term4 part 2 robust}
\end{align}
By using \eqref{c5 term 4 part 1 robust} and \eqref{c5 term4 part 2 robust} in \eqref{c5 term4 complete} and then substituting \eqref{c5 term1} \eqref{c5 term 2 robust}, \eqref{c5 term 3 robust}, and \eqref{c5 term4 complete} in \eqref{mean square error terms for RS APA Robust} we can calculate the MSE, which is given by \eqref{mean square error APA robust}. This concludes the proof. \vspace{-1.5em}
\section{Proof of the MSE for the APA}\label{Appendix MSE APA}
Here, we describe in detail how to obtain the MSE employed in the APA algorithm. Let us first consider the MSE, which is given by
\begin{align}
    \mathbb{E}\left[\varepsilon\right]=&{\mathbb{E}\left[\mathbf{s}^{\left(\text{RS}\right)^H}\mathbf{s}^{\left(\text{RS}\right)}\right]}-{\mathbb{E}\left[\mathbf{s}^{\left(\text{RS}\right)^H}\mathbf{y}'\right]}-{\mathbb{E}\left[\mathbf{y'}^H\mathbf{s}^{\left(\text{RS}\right)}\right]}+{\mathbb{E}\left[\mathbf{y'}^H\mathbf{y}'\right]}.\label{mean square error terms for RS APA}
\end{align}
The first term of \eqref{mean square error terms for RS APA} is computed identically to \eqref{c5 term1}.

By taking the expected value of the second term in \eqref{mean square error terms for RS APA} and expanding the equation, we have
\begin{align}
\mathbb{E}\left[\mathbf{s}^{\left(\text{RS}\right)^H}\mathbf{y}'\right]=&a_c\mathbb{E}\left[s_c^*s_c\right]\sum_{l=1}^M\mathbf{h}_{l,*}\mathbf{p}_c+\sum_{l=1}^M \mathbb{E}\left[s_c^*n_l\right]+ \sum_{l=1}^{M}\mathbf{h}_{l,*}\sum_{j=1}^{M}a_j \mathbb{E}\left[s_c^*s_j\right] \mathbf{p}_j\nonumber\\
&+a_c\sum_{l=1}^M\mathbb{E}\left[s_l^*s_c\right]\mathbf{h}_{l,*}\mathbf{p}_c+\sum_{q=1}^M\sum_{l=1}^{M}a_l \mathbb{E}\left[s_q^*s_l\right]\mathbf{h}_{q,*}\mathbf{p}_l+\sum_{l=1}^M\mathbb{E}\left[s_l^*n_l\right].\label{c5 term 2 full}
\end{align}
Since the symbols are uncorrelated, equation \eqref{c5 term 2 full} is reduced to
\begin{align}
    \mathbb{E}\left[\mathbf{s}^{\left(\text{RS}\right)^H}\mathbf{y}'\right]=a_c\sum_{j=1}^{M}\mathbf{h}_{j,*}\mathbf{p}_c+\sum_{l=1}^M a_l \mathbf{h}_{l,*}\mathbf{p}_l.\label{c5 term2}
\end{align}
The third term of \eqref{mean square error terms for RS APA} can be computed in a similar way as the second term, which lead us to
\begin{align}
    \mathbb{E}\left[\mathbf{y'}^H\mathbf{s}^{\left(\text{RS}\right)}\right]=a_c\sum_{j=1}^{M}\left(\mathbf{h}_{l,*}\mathbf{p}_c\right)^{*}+\sum_{l=1}^M a_l \left(\mathbf{h}_{l,*}\mathbf{p}_l\right)^{*}.\label{c5 term3}
\end{align}

The last term of \eqref{mean square error terms for RS APA} is equal to
\begin{align}
    \mathbb{E}\left[ \mathbf{y'}^H\mathbf{y}'\right]=\sum_{l=1}^M\sum\limits_{\substack{j=1\\j\neq l}}^M\mathbb{E}\left[y_l^* y_j\right]+2\sum_{i=1}^{M}\mathbb{E}\left[y_i^* y_i\right],\label{c5 term4 APA complete}
\end{align}
Let us first compute the quantity given by
\begin{equation}
    \mathbb{E}\left[y_i^* y_i\right]=a_c^2\lvert\mathbf{h}_{i,*}\mathbf{p}_c\rvert^2+\sum_{l=1}^M a_l^2\lvert\mathbf{h}_{i,*}\mathbf{p}_l\rvert^2+\sigma_n^2.\label{c5 term 1 of term4}
\end{equation}
Additionally, we know that
\begin{align}
    \mathbb{E}\left[y_i^* y_j\right]=&\mathbb{E}\left[\left(a_c s_c \mathbf{h}_{i,*}\mathbf{p}_c+\sum_{q=1}^{M}a_q s_q \mathbf{h}_{i,*}\mathbf{p}_q+ n_i\right)^*\right.\times\left.\left(a_c s_c \mathbf{h}_{j,*}\mathbf{p}_c+\sum_{l=1}^{M}a_l s_l \mathbf{h}_{j,*}\mathbf{p}_l+ n_j\right)\right],\nonumber\\
    =&a_c^2\left(\mathbf{h}_{i,*}\mathbf{p}_c\right)^*\left(\mathbf{h}_{j,*}\mathbf{p}_c\right)+\sum_{l=1}^Ma_l^2\left(\mathbf{h}_{i,*}\mathbf{p}_l\right)^*\left(\mathbf{h}_{j,*}\mathbf{p}_l\right),\label{c5 term 2 of term4}
\end{align}
for all $i\neq j$. From the last equation, we have
\begin{align}
    \sum_{i=1}^M\sum\limits_{\substack{j=1\\j\neq l}}^M\mathbb{E}\left[y_l^* y_j\right]=&\sum_{i=1}^M\sum\limits_{\substack{j=1\\j\neq l}}^M\left[a_c^2\left(\mathbf{h}_{i,*}\mathbf{p}_c\right)^*\left(\mathbf{h}_{j,*}\mathbf{p}_c\right)+\sum_{l=1}^Ma_l^2\left(\mathbf{h}_{i,*}\mathbf{p}_l\right)^*\left(\mathbf{h}_{j,*}\mathbf{p}_l\right)\right],\nonumber\\
    =&\sum_{i=1}^M\sum\limits_{\substack{j=1\\j\neq l}}^Ma_c^2\left(\mathbf{h}_{i,*}\mathbf{p}_c\right)^*\left(\mathbf{h}_{j,*}\mathbf{p}_c\right)+\sum_{i=1}^M\sum\limits_{\substack{j=1\\j\neq l}}^M\sum_{l=1}^Ma_l^2\left(\mathbf{h}_{i,*}\mathbf{p}_l\right)^*\left(\mathbf{h}_{j,*}\mathbf{p}_l\right),\nonumber\\
    =&\sum_{i=1}^M\sum\limits_{\substack{j=1\\j\neq l}}^Ma_c^2\phi^{\left(i,c\right)^*}\phi^{\left(j,c\right)}+\sum_{i=1}^M\sum\limits_{\substack{j=1\\j\neq l}}^M\sum_{l=1}^Ma_l^2\phi^{\left(i,l\right)^*}\phi^{\left(j,l\right)},\label{c5 term 2 of term4 extended}
\end{align}
where we define $\phi^{\left(i,c\right)}=\mathbf{h}_{l,*}\mathbf{p}_c$ and $\phi^{\left(i,l\right)}=\mathbf{h}_{i,*}\mathbf{p}_l$ for all $i,l \in \left[1,M\right]$. Applying the property $\phi^{\left(i,l\right)^*}\phi^{\left(j,l\right)}+\phi^{\left(i,l\right)}\phi^{\left(j,l\right)^*}=2\Re\left[\phi^{\left(i,l\right)^*}\phi^{\left(j,l\right)}\right]$, we can simplify half of the sums from the triple summation, i.e.,
\begin{equation}
\sum_{i=1}^M\sum\limits_{\substack{j=1\\j\neq l}}^M\sum_{l=1}^Ma_l^2\phi^{\left(i,l\right)^*}\phi^{\left(j,l\right)}=2\sum_{i=1}^{M-1}\sum_{q=i+1}^{M}\sum_{r=1}^M a_r^2\Re\left[\phi^{\left(i,r\right)^*}\phi^{\left(q,r\right)}\right].\label{c5 triple summation explained}
\end{equation}

The final step to obtain the last term of \eqref{mean square error terms for RS APA} is to employ \eqref{c5 term 1 of term4}, \eqref{c5 term 2 of term4 extended}, and \eqref{c5 triple summation explained} to compute the following quantities:
\begin{equation}
\sum_{i=1}^{M}\mathbb{E}\left[y_i^* y_i\right]=\sum_{l=1}^M a_c^2\lvert\mathbf{h}_{l,*}\mathbf{p}_c\rvert^2+\sum_{i=1}^M\sum_{j=1}^M a_j^2\lvert\mathbf{h}_{i,*}\mathbf{p}_j\rvert^2+M\sigma_n^2,\label{c5 term4 a}
\end{equation}

\begin{align}
    \sum_{l=1}^M\sum\limits_{\substack{j=1\\j\neq l}}^M\mathbb{E}\left[y_l^* y_j\right]=&2\sum_{i=1}^{M-1}\sum_{q=i+1}^{M}\sum_{r=1}^M a_r^2\Re\left[\phi^{\left(i,r\right)^*}\phi^{\left(q,r\right)}\right]+\sum_{l=1}^M\sum\limits_{\substack{j=1\\j\neq l}}^M a_c^2\phi^{\left(l,c\right)^*}\phi^{\left(j,c\right)}.\label{c5 term4}
\end{align}

 By using \eqref{c5 term4 a} and \eqref{c5 term4} in \eqref{c5 term4 APA complete}  and then substituting \eqref{c5 term1}, \eqref{c5 term2}, \eqref{c5 term3}, \eqref{c5 term4 APA complete}   in \eqref{mean square error terms for RS APA} we get the MSE in \eqref{mean square error APA RS}. \vspace{-1.5em}




\ifCLASSOPTIONcaptionsoff
  \newpage
\fi



%



\bibliographystyle{IEEEtran}
\bibliography{tiny}


%








\end{document}